\documentclass{aastex631}
\usepackage{bm}
\usepackage{subcaption}
\usepackage[T1]{fontenc}    
\usepackage{booktabs}       
\usepackage{threeparttable}
\usepackage{multirow}

\usepackage{amsmath}

\begin{document}

 \title{Exploring crowd persistent dynamism from pedestrian crossing
perspective: An empirical study}

\author{Jinghui Wang}
\affiliation{School of Safety Science and Emergency Management\\
Wuhan University of Technology\\
Wuhan, China}

\author{Wei Lv}
\altaffiliation{{\url{weil@whut.edu.cn}}}
\affiliation{School of Safety Science and Emergency Management\\
Wuhan University of Technology\\
Wuhan, China}

\author{Huihua Jiang}
\affiliation{School of Safety Science and Emergency Management\\
Wuhan University of Technology\\
Wuhan, China}

\author{Zhiming Fang}

\affiliation{Business School\\
University of Shanghai for Science and Technology\\
Shanghai, China}

\author{Jian Ma}
\affiliation{School of Transportation and Logistics\\
Southwest Jiaotong University\\
Chengdu, China}



\begin{abstract}

Crowd studies have gained increasing relevance due to the recurring incidents of crowd crush
accidents. In addressing the issue of the crowd’s persistent dynamism, this paper explored the
macroscopic and microscopic features of pedestrians crossing in static and dynamic contexts,
employing a series of systematic experiments. Firstly, empirical evidence has confirmed the existence
of crowd’s persistent dynamism. Subsequently, the research delves into two aspects,
qualitative and quantitative, to address the following questions: (1) Cross pedestrians tend to
avoid high-density areas when crossing static crowds and particularly evade pedestrians in front
to avoid deceleration, thus inducing the formation of cross-channels, a self-organization phenomenon.
(2) In dynamic crowds, when pedestrian suffers spatial constrained, two patterns
emerge: decelerate or detour. Research results indicate the differences in pedestrian crossing
behaviors between static and dynamic crowds, such as the formation of crossing channels,
backward detours, and spiral turning. However, the strategy of pedestrian crossing remains
consistent: utilizing detours to overcome spatial constraints. Finally, the empirical results of this
study address the final question: pedestrians detouring causes crowds’ persistent collective
dynamism. These findings contribute to an enhanced understanding of pedestrian dynamics in
extreme conditions and provide empirical support for research on individual movement patterns
and crowd behavior prediction.

\end{abstract}



\section{Introduction} 
\label{section1}

United Nations projections anticipate a global population surge to 9.8 billion by 2050, with nearly 7 billion inhabiting urban areas
due to accelerating urbanization \citep{nations2015world}. The presence of dense human populations engenders challenges in crowd
management and leads to annual occurrences of stampede accidents arising from crowd congregations globally \citep{feliciani2023trends}.

Extensive studies indicate that ultra-high crowd density constitutes merely one facet in the sequence of events leading to stampedes.
Intriguingly, despite frequent overcrowding in subway cars worldwide, crowd crush accidents have never been reported. One
contributing factor is the spatial constraints inherent in confined subway cars, which effectively shield pedestrians from experiencing
crushing forces. Furthermore, the lack of motivation for movement among subway crowds, typically stationary, establishes a stable physical structure. Almost all reports on crowd crush accidents consistently observe and document that the entire crowd is in dynamism
prior to the stampede incident, a viewpoint substantiated by field research \citep{helbing2007dynamics} . However, does the dynamic
nature of a crowd inevitably lead to crowd crush accidents? Observation of nature addressed this question. Clustering phenomena are
quite common in the biological realm, yet crowd crushes are rarely observed in natural animal groups, such as bird flocks and fish
schools \citep{couzin2018synchronization}. Criticality plays a role in making them immune to crowd crush phenomena. Near the critical point,
remarkable properties emerge spontaneously, where individuals’ behavior becomes correlated regardless of their distance from each
other. This correlation is mathematically equivalent to information propagation with minimal loss over the entire structure. A study
discovered that a flock of hundreds of starlings can collectively change direction in nearly half a second \citep{attanasi2014information}. In
contrast, human control and response to movement are significantly delayed. In this regard, the terrestrial organisms of sheep herds \citep{pastor2015experimental}, mice \citep{saloma2015prior}, ants \citep{wang2015behavior}, along with granular media \citep{zuriguel2011silo,patterson2017clogging} are more similar to the human crowd. Therefore, some believe in human crowds, the “anisotropy” of collective dynamics
induces shock waves, resulting in stampede accidents \citep{zhang2012fundamental,bottinelli2016emergent,feliciani2018measurement,zanlungo2023pure}.

Our understanding of crowd remains unclear. In response to these challenges, a series of supervised pedestrian crossing experiments
were conducted in this paper. The findings will provide new insights into understanding and predicting crowd dynamic. The
manuscript is structured as follows: an overview of past research on related topics is provided in the next section, followed by presenting
the question addressed by this paper. Section \ref{section3} delineates the corresponding experimental procedures and data processing
methodologies. Section \ref{section4} gives a comprehensive introduction to the pertinent observable variables, and the examination of the existence
of the question is presented in Section \ref{section5}. Subsequently, Sections \ref{section6} and \ref{section7} offer a thorough analysis of pedestrian detour
mechanisms by amalgamating macroscopic and microscopic approaches and combining qualitative and quantitative analyses. Finally,
Section \ref{section8} encapsulates the conclusions derived from the study.

\section{Literature review} \label{section2}

Delivering a comprehensive review of advancements in human behavioral dynamics, particularly dynamic crowds, is crucial to
understanding the current challenges and identifying areas requiring further exploration and attention. Empirical methodologies of
crowd dynamic research encompass two primary strategies: field investigations \citep{haghani2020f} and supervised experiments \citep{haghani2020e}.

In field methods, researchers have predominantly relied on video data obtained from Closed-Circuit Television (CCTV) for their
investigations, enabling the distinction between normal behavior and emergency behavior. Studies focused on normal behavior have
commonly explored settings such as subway stations \citep{zanlungo2014potential,zuriguel2020contact}, crowd gatherings during festivals \citep{zhang2013empirical}, and crosswalks \citep{zeng2017specification}, among others. However, field studies concerning emergency behavior have
garnered greater interest among researchers, as it often deviates from the typical behavior patterns. Examples of such emergencies
include turbulence in dense crowds \citep{helbing2007dynamics}, the fundamental diagram area prior to pedestrians falling \citep{parisi2021pedestrian}, and various instances like earthquake scenarios \citep{li2015parameter}, the Hajj pilgrimage \citep{helbing2007dynamics}, the love parade
disaster \citep{pretorius2015large,ma2013new}, the bull running festival \citep{parisi2021pedestrian}, and terrorist attacks \citep{wang2019empirical}, to name a few.

Supervised experiments involve recruiting participants to conduct pedestrian dynamics experiments in predefined scenarios. A
well-lit environment and accurate coordinate system calibration can significantly enhance data precision for micro-level analysis. This
approach presents a broad spectrum of research potentialities and is empowered to conduct nuanced individual-level evaluations.
Implementing supervised experiments with human participants has seen a notable upswing in recent years, primarily due to their
utility in studying crowd behavior. Through setting up predefined regions, supervised experiments have facilitated the observation of
pedestrian behavior in various situations, such as uni-directional flow \citep{fu2021effect,fujita2019traffic}, bi-directional flow \citep{zhang2012fundamental,jin2019observational}, cross-flow \citep{cao2017fundamental,yamamoto2019body}, and bottleneck situations \citep{li2021comparative,garcimartin2018redefining,zhao2017self}, group behavior \citep{hu2020experimental,ma2017experimental}, vortex motion \citep{echeverria2022spontaneous},
avoidance \citep{parisi2016experimental,moussaid2009experimental}, intrusion \citep{nicolas2019mechanical,kleinmeier2020agent}, etc. These studies provide
extensive fundamental diagram patterns and statistics of pedestrian movement characteristics, such as pedestrian walking preference \citep{moussaid2009experimental}, relaxation time \citep{ma2010experimental}, ’V’-like and ’U’-like formation of the pedestrian social group \citep{moussaid2010walking,zanlungo2014potential,zanlungo2015spatial}, synchronized flow \citep{ma2017experimental}, etc.

The bi-directional flow scenario has garnered significant attention in pedestrian experiments, with researchers exploring lane
formation and the underlying mechanisms of pedestrian movement. When pedestrian crowds in opposite directions encounter each
other, such as at a pedestrian crosswalk, they initially form small channels that merge to create wider lanes \citep{helbing2005self}. This
phenomenon is associated with fluctuation-induced ordering, which reduces the number of interfaces where “friction” occurs between
opposing directions of motion. Similar to the traffic system, smaller congested groups gradually converge to form larger moving jams.
This self-organizing process tends to steer the traffic system towards a more stable physical structure. However, \citet{moussaid2012traffic} 
discovered structural instabilities in lane formation. Slower walkers create density gaps, while faster walkers exploit these gaps to
overtake pedestrians ahead of them. These local interactions lead to large-scale traffic breakdowns, and the spontaneous self-organization
results in a sub-optimal state. Experiments conducted within circular channels with periodic boundary conditions
have shed light on the continuous evolution of pedestrian flow. Notably, significant differences in step length and stride frequency have
been observed between single-file pedestrians on the inner and outer sides of the circular passage \citep{jelic2012steping}, although these
differences may not have a decisive impact. Given the distinct circumferences of the inner and outer circles of the circular passage,pedestrians tend to walk along the inner side. This preference makes it challenging for pedestrians to stagger when conflicts arise,
potentially contributing to the breakdown of the lane.

Characterizing the dynamics of crowds in high-density scenarios through supervised experiments poses a challenging endeavor.
Nevertheless, the utilization of bottleneck and cross-flow experiments stands out as a highly effective approach in this realm. Pedestrians
were found to tend to decrease their speed in merging regions, resulting in a notable reduction in flow rate, especially when
dealing with larger merging angles \citep{shiwakoti2015examining}. The significance of this observation lies in the need to consider architectural
design factors when comprehending and optimizing pedestrian movement within urban environments. Distinct neighbor
distribution characteristics for three types of pedestrian movements, namely same direction (U-ped), opposite direction (B-ped), or
cross direction (C-ped), were identified through research on pedestrian spatial behavior \citep{cao2021spatial}. Notably, diversification in
flow direction enhances the dynamism within crowds under conditions of high density, as demonstrated in fundamental diagram
research \citep{feliciani2022introduction}. Although the phenomenon of self-organized interference stripes formed by crosswalk pedestrian flows
is well-known, however, further work and exploration are needed to conduct additional quantitative measurements and evaluations in
this regard \citep{zanlungo2023stripe,mullick2022analysis}. Furthermore, the research on individuals crossing crowds, particularly in a
static context, emphasizes the distinctions between crowd and granular media. Pedestrian movements are observed to be influenced by
psychological and social norms, which are challenging to express through mechanical descriptions \citep{nicolas2019mechanical,kleinmeier2020agent}. What’s more, the controversial topic of crowd dynamics was reintroduced into the discussion, e.g., can social groups lead
to an increase or decrease in the evacuation time of a crowd \citep{bode2015disentangling,von2017empirical}? Could obstacles be
beneficial for pedestrian evacuation \citep{zuriguel2020contact,garcimartin2018redefining,frank2011room,shi2019examining}? Table \ref{table1} details
the supervised pedestrian experiments studies discussed above in recent years and classified based on crowd dynamic characteristics.

From Table \ref{table1} and relevant statistics \citep{haghani2018crowd}, it is evident that research on uni and bi-directional flow still dominates.
Comparatively, studies on multi-directional conflict processes and other frequently occurring but overlooked behaviors, such as
steering and avoidance in crowds, have been marginalized. Controlled experimental studies encounter inherent constraints in
analyzing large-scale crowds due to ethical issues and safety considerations. Experiments involving extremely high-density crowds are
impracticable, making phenomena like stampedes and shockwaves highly improbable to be observed within the scope of controlled
experiments. To overcome these limitations, the application of Virtual Reality (VR) technology has emerged as a promising tool for
quantitatively evaluating the individual behavior of large-scale crowds \citep{zhao2020assessing,tong2023investigation}.

Pioneering investigations have dedicated substantial effort toward elucidating the movement mechanisms of crowds through
empirical analysis. Prevalent theories suggest the presence of turbulence or shockwaves within dense crowds as antecedents to crowd
crush incidents, thereby accentuating the criticality of acquiring an exhaustive understanding of crowd turbulence phenomena. Specifically, in the context of ultra-high-density crowds, the limited space availability between individuals exerts significant constraints
on their mobility. Yet, intriguingly, the collective behaviour of the crowd remains dynamic, which posits an interesting
question: Why does the collective dynamism of the crowd persist despite these physical limitations? This paper aims to address this
gap.

\begin{longrotatetable}

\begin{table}[!ht]
\centering
\caption{Studies based on supervised pedestrian experiments (M: male, F: female).}
\begin{threeparttable}

\begin{tabular}{c}
\hline
\includegraphics[height=16cm]{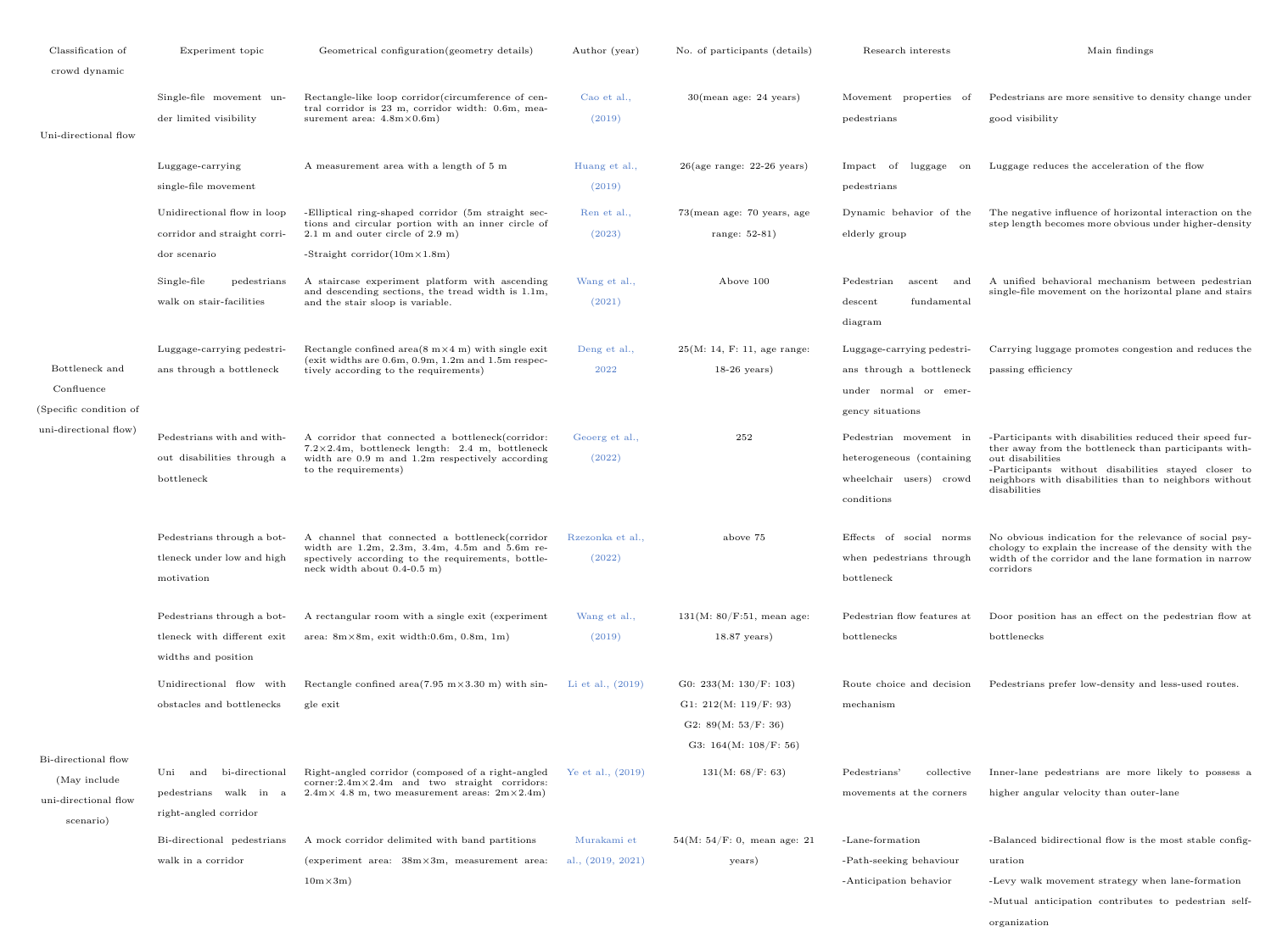}\\

\hline
\end{tabular}

 \begin{tablenotes}
        \footnotesize
        \item Note: Due to length limitations and potential oversights, some homogenized studies, as well as certain crucial but overlooked research,
were not included in the table.
\item Relevant papers: \citep{cao2019dynamic,huang2019experimental,ren2023comparison,wang2021pedestrian,deng2022effect,geoerg2022people,rzezonka2022attempt,wang2019empirical,li2019comparing,ye2019experimental}

      \end{tablenotes}
    \end{threeparttable}
\label{table1}
\end{table}

\begin{table}[!ht]\ContinuedFloat
\centering
\caption{(Continue).}
\begin{threeparttable}

\begin{tabular}{c}
\hline
\includegraphics[height=16cm]{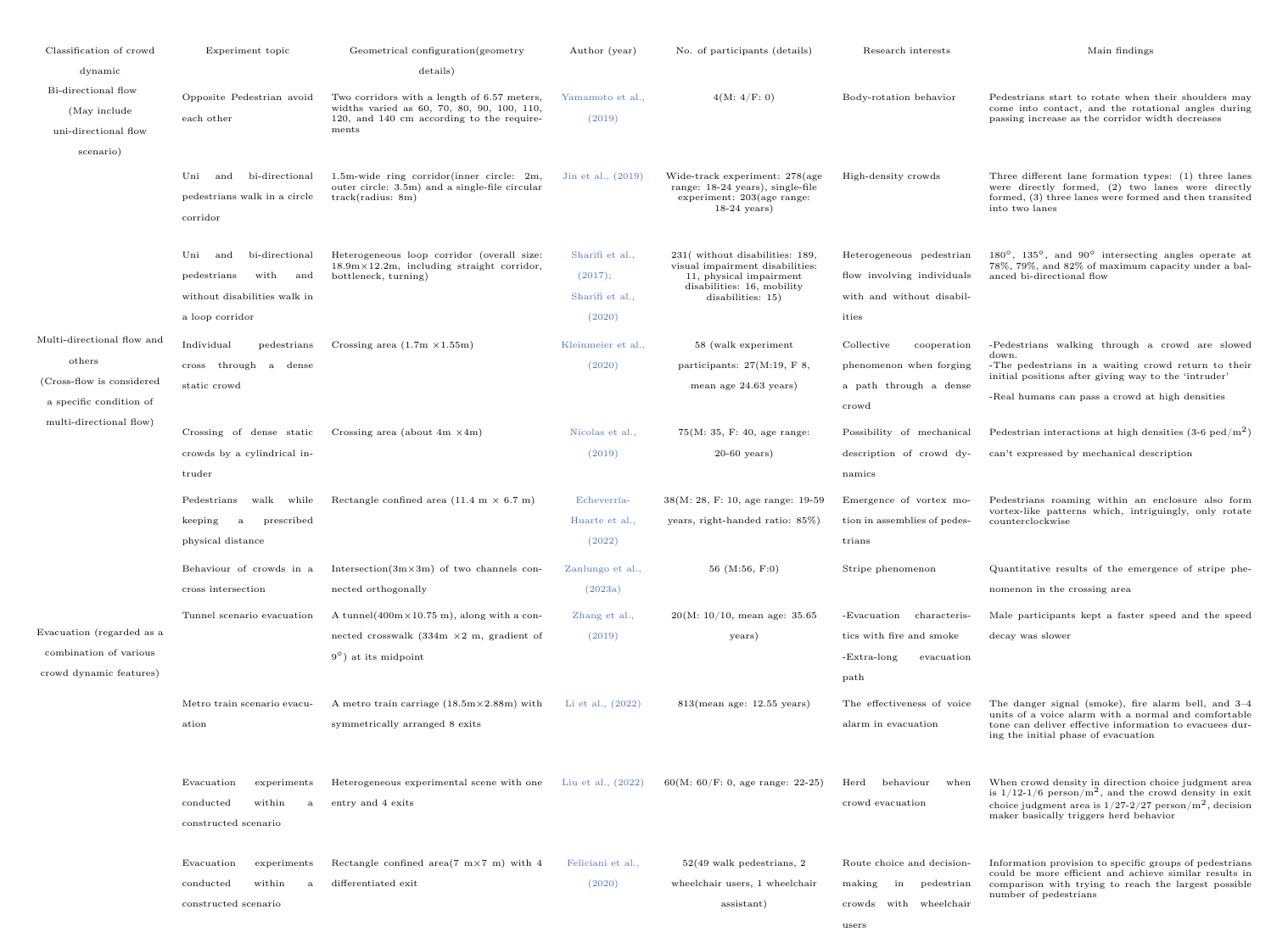}\\

\hline
\end{tabular}

 \begin{tablenotes}
        \footnotesize
        \item Relevant papers: \citep{murakami2019levy,murakami2021mutual,yamamoto2019body,jin2019observational,sharifi2017large,sharifi2020exploring,kleinmeier2020agent,nicolas2019mechanical,echeverria2022spontaneous,zanlungo2023stripe,zhang2019experimental,li2022experimental,liu2022experimental,feliciani2020efficiently}

      \end{tablenotes}
    \end{threeparttable}
    \label{table1}
\end{table}
\end{longrotatetable}

\section{Experiments} \label{section3}

\subsection{Background} \label{subsection3.1}

\begin{figure}[ht!]
\plotone{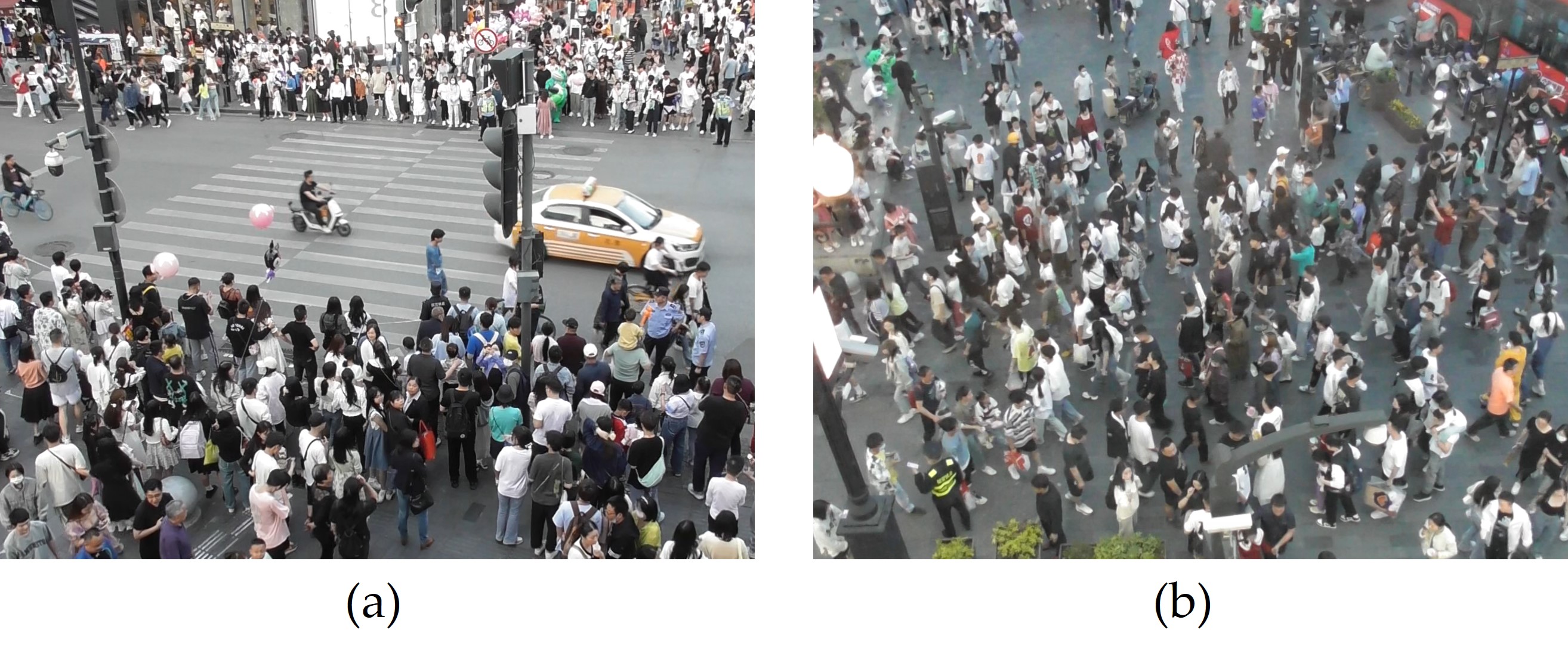}
\caption{The static and dynamic crowds in public space, (a) The static crowd awaiting at the crossroad, (b) The dynamic crowd on the T-shaped
street. Captured on May 1, 2023, International Workers’ Day. Location: Jianghan Road, Wuhan City. 
\label{fig1}}
\end{figure}

Static and dynamic crowds represent two standard crowd states. Static crowds refer to stationary conditions such as pedestrians
awaiting traffic signals, audiences at music festivals, and individuals queuing at service windows. Conversely, dynamic crowds are
frequently observed in areas like streets, squares, and stations during holidays, as demonstrated in Fig.\ref{fig1}. In all social activities
involving human participation, the phenomenon of crossing is virtually ubiquitous. For instance, security personnel at concerts or
traffic police officers often cross through crowds, and commuters travel crowded squares, tourists during holidays, etc.
Considering pedestrians’ crossing behavior in both contexts, corresponding experiments have been conducted. Investigation of
pedestrians’ crossing character aims to address the questions raised in this paper.

\subsection{Experimental setup}\label{subsection3.2}

Experiments were undertaken at the Wuhan University of Technology campus in China on May 1, 2019. The study engaged 50
participants, a blend of undergraduate and master students aged between 19 and 22, maintaining an equal gender distribution.

\begin{figure}[ht!]

\centering         
\includegraphics[scale=0.5]{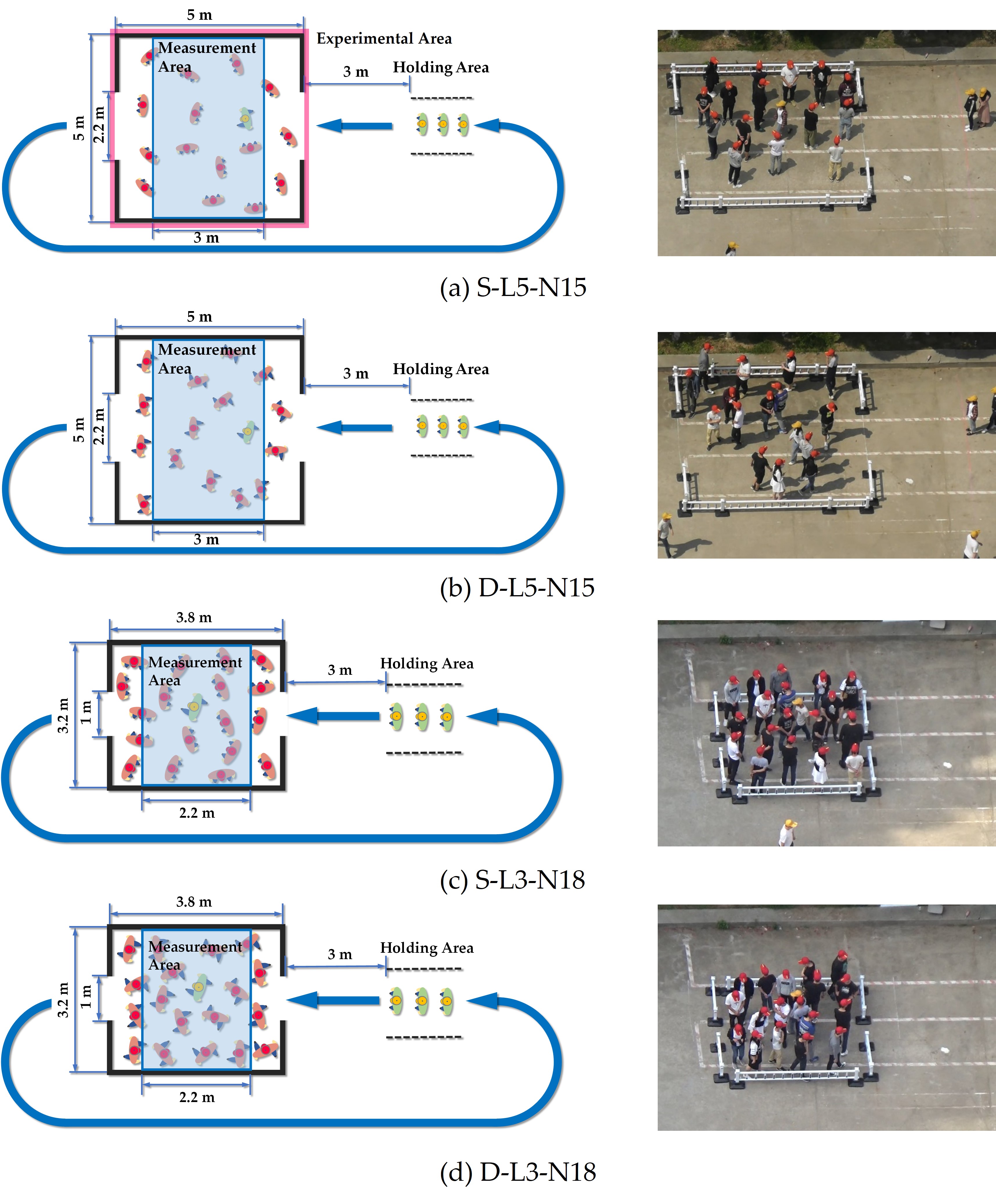}
\caption{Experimental scenarios and schematics (Left: schematic diagram of the different crossing experiments, right: experiment snapshots. Subfigures
(a) and (b) represent high-density experiments, and subfigures (c) and (d) represent low-density experiments). (A more detailed explanation
is warranted, S-L5-N15: In the experimental area measuring 5 m by 5 m, 15 volunteers were randomly positioned. The volunteers remained stationary
throughout the experiment while the crossing pedestrians waited in the holding area. Upon hearing the instructions, they entered the
entrance of the experimental area and exited through the designated exit, signifying the completion of a single cross-experiment). 
\label{fig2}}
\end{figure}

In the experimental design, an individual was assigned to cross a predetermined experimental zone that accommodated a varying
number of other participants. Two primary control conditions were established: global density and motion patterns (comprising static
and dynamic). The static pattern required the participants in the experimental zone to maintain a stationary position, whereas the
dynamic pattern necessitated participants engage in spontaneous movement during the experiment. These experiments were classified
into low-density and high-density scenarios based on the dimensions of the experimental zones. The low-density experiment utilized a
rectangular zone of 5 m × 5 m, whereas the high-density experiment employed a rectangular zone measuring 3.8 m × 3.2 m. A
schematic illustration of the experimental settings is presented in Fig.\ref{fig2}, offering a comparative visualization of varying crowd patterns
and the experimental conditions in both high-density and low-density scenarios.

Furthermore, a prudent approach was adopted to circumvent potential inaccuracies in the estimation of individual local density
when participants were close to the boundaries of the experimental zone. Thus, we selected only the central three-fifths portion of the
experimental zone for our statistical analysis as the designated measurement area. A schematic diagram demonstrating this selected
measurement area is depicted in Fig.\ref{fig2}.

\subsection{Experimental procedure\label{subsection3.3}}


During the participant’s recruitment phase, the personal characteristics: height, and gender were meticulously documented. Before
initiating the experiment, participants were briefed about the experimental protocol and issued corresponding hats. As the experiment
commenced, the crossing pedestrians assembled in a designated waiting area. Upon receiving the starting instruction, the crossing
pedestrians performed the crossing action sequentially. Subsequent pedestrians initiated their crossing only after the preceding pedestrian had passed the experimental zone. Communication among participants was expressly prohibited throughout the experiment’s
duration. The experiment for a specific group was concluded once the predetermined number of repetitions was met. The
repetition count for each experimental group is outlined in Table \ref{table2}.

All participants in the experiment were purposefully balanced to a mixed-gender composition. This intentional selection aimed to
harmonize the experiment with the gender characteristics common in everyday social clusters. For an exhaustive overview of the
experimental setup, please refer to Table \ref{table2}.

\begin{table}[]
\caption{Setup for crossing experiment.}
\centering
\resizebox{.9\columnwidth}{!}{%
\centering
\begin{tabular}{lllllll}
	\toprule
Index &
  Way of movement &
  Geometrical configuration   (m2) &
  Crossing distance (m) &
  No. of participants &
  \begin{tabular}[c]{@{}l@{}}Global density\\    \\ (ped/m2)\end{tabular} &
  No. of repetitions \\ \midrule
S(D)-L5-N0  & Static (Dynamic) & 5×5     & 5   & 0  & 0    & 30      \\
S(D)-L5-N5  & Static (Dynamic) & 5×5     & 5   & 5  & 0.2  & 50 (39) \\
S(D)-L5-N10 & Static (Dynamic) & 5×5     & 5   & 10 & 0.4  & 50 (39) \\
S(D)-L5-N15 & Static (Dynamic) & 5×5     & 5   & 15 & 0.6  & 49 (39) \\
S(D)-L5-N20 & Static (Dynamic) & 5×5     & 5   & 20 & 0.8  & 50 (40) \\
S(D)-L5-N25 & Static (Dynamic) & 5×5     & 5   & 25 & 1    & 49 (38) \\
S(D)-L5-N30 & Static (Dynamic) & 5×5     & 5   & 30 & 1.2  & 50 (39) \\
S(D)-L5-N35 & Static (Dynamic) & 5×5     & 5   & 35 & 1.4  & 49 (34) \\
S(D)-L5-N40 & Static (Dynamic) & 5×5     & 5   & 40 & 1.6  & 45 (35) \\
S(D)-L5-N45 & Static (Dynamic) & 5×5     & 5   & 45 & 1.8  & 51 (23) \\
S(D)-L5-N49 & Static (Dynamic) & 5×5     & 5   & 49 & 1.96 & 24 (20) \\
S(D)-L3-N18 & Static (Dynamic) & 3.8×3.2 & 3.8 & 18 & 1.48 & 29 (30) \\
S(D)-L3-N24 & Static (Dynamic) & 3.8×3.2 & 3.8 & 24 & 1.97 & 29 (29) \\
S(D)-L3-N30 & Static (Dynamic) & 3.8×3.2 & 3.8 & 30 & 2.47 & 29 (29) \\
S(D)-L3-N36 & Static (Dynamic) & 3.8×3.2 & 3.8 & 36 & 2.96 & 29 (29) \\
S(D)-L3-N42 & Static (Dynamic) & 3.8×3.2 & 3.8 & 42 & 3.45 & 16 (17) \\
\bottomrule
\end{tabular}%
}
\label{table2}
\end{table}

Two high-definition cameras (1920 × 1080 pixel resolution, 25 fps) were positioned atop a nearby building, at an approximate
height of 100 m, to document the entire experimental process comprehensively. The trajectories were manually extracted and then
transformed into physical coordinates using the Direct Linear Transformation (DLT) method \citep{hartley2003multiple}. The approximate
trajectory representation was achieved by tracking the head positions of individuals across each video frame. Notably, potential errors emanating from height variations among the participants when mapping pixel coordinates to real-space coordinates were intentionally
overlooked. Therefore, during the trajectory extraction process, trajectory data were segregated based on the gender of volunteers to
mitigate the potential influence of height variations, thereby ensuring that the error was controlled within a 10 cm margin.

\section{Observable variables\label{section4}}
\begin{figure}[ht!]
\centering
\includegraphics[scale=0.3]{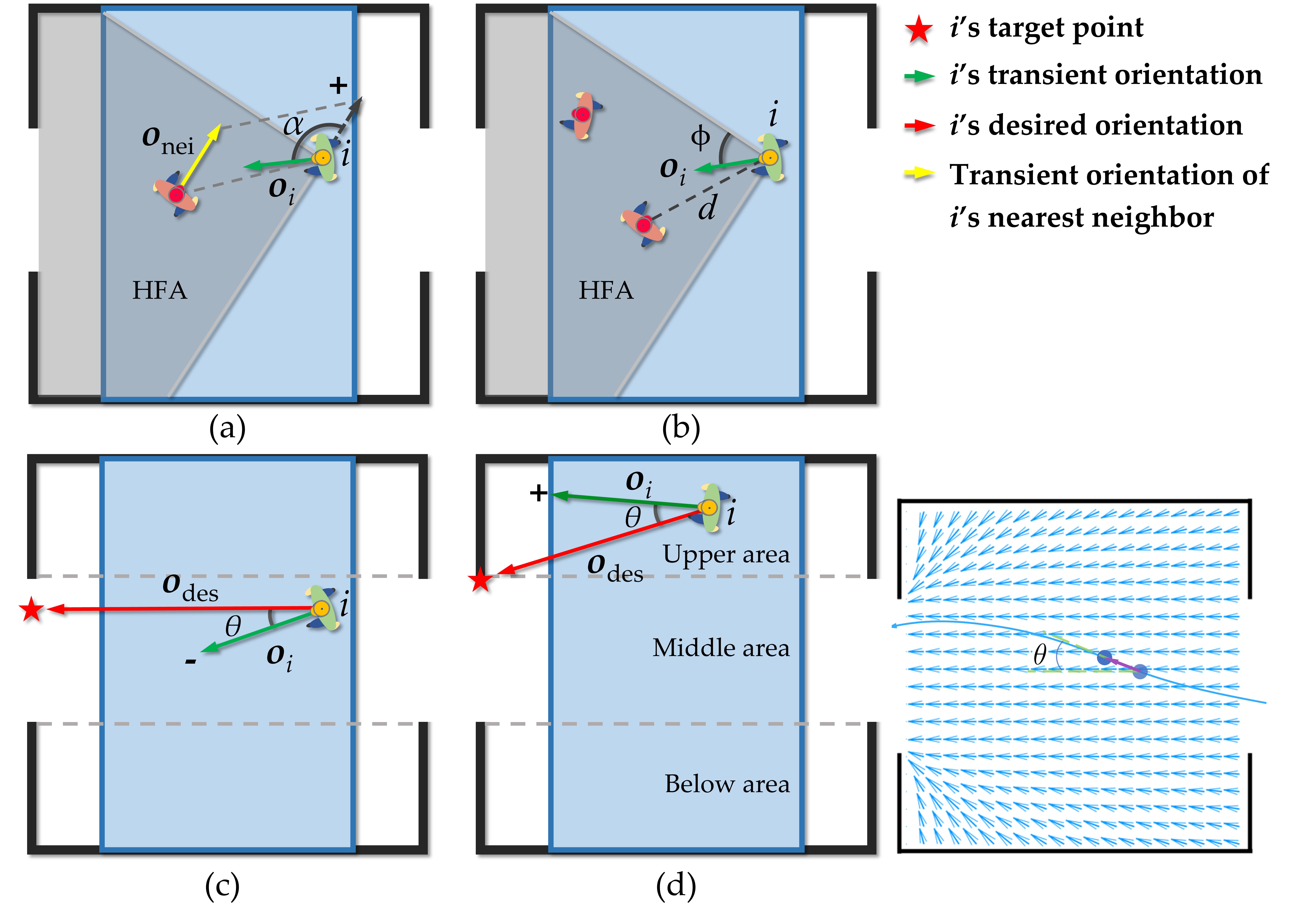}
\caption{Schematic diagram of the relevant observable variables of cross pedestrian in the experiment. (clockwise as a positive value, counterclockwise
as a negative value). 
\label{fig3}}
\end{figure}

In our research, we employ a comprehensive set of 9 observable variables to quantify the dynamics of cross phenomena accurately.
The corresponding schematic representation is presented in Fig.\ref{fig3}. This section is dedicated to introducing these observable variables
(see Table \ref{table3}), accompanied by a quantitative description that elucidates their precise definitions and inherent significance.

\begin{table}[!htb]\centering
\caption{Summary of observable variables.}
\resizebox{.5\columnwidth}{!}{
\begin{tabular}{lll}
\toprule
Name                              & Symbol & Definition \\ \midrule
Instantaneous velocity            & $v$      & Eq.(\ref{1})     \\
Local density                     & $\rho$      & Eq.(\ref{2})    \\
Flow rate                         & $J$      & Eq.(\ref{3})    \\
Eccentricity attentional   angle. & $\phi$      & /          \\
Deviation angle                   & $\theta$      & Eq.(\ref{4})    \\
NNRD                              & $d$      & Eq.(\ref{5})    \\
NNRA                              & $\alpha$      & Eq.(\ref{6})    \\
Angular velocity                  & $\omega$      & Eq.(\ref{7})    \\
\bottomrule
\end{tabular}
}
\label{table3}
\end{table}

The observable variables include:

• Instantaneous velocity

• Global density

• Local density

• Flow rate

• Eccentricity attentional angle

• Deviation angle

• Nearest Neighbor Relative Distance (NNRD)

• Nearest Neighbor Relative Angle (NNRA)

• Angular velocity.

\centerline{}

1 Instantaneous velocity ($v$)

The instantaneous velocity denotes the average speed of an individual pedestrian within a sampling time interval $\tau$. Due to the short
duration of the sampling time interval, the velocity sampling can be regarded as transient. The formula for calculating the instantaneous velocity of a pedestrian at time t is provided by Eq.(\ref{1}).

\begin{equation}\label{1}
{{v}_{t}}=\frac{\left\| {\bm{x}_{t+\tau /2}}-{\bm{x}_{t-\tau /2}} \right\|}{\tau }
\end{equation}

where $ \bm{x}_{t}$ denotes the coordinates of the pedestrian at the moment $t$, $\tau$ represents the sampling time gap. Velocity sampling was performed
every 10 frames ($\tau$ = 0.4 s).

\centerline{}

2 Global density

The global density refers to the average pedestrian density within the experimental area, which characterizes the overall distribution
of the crowd within the experimental region. It is calculated as the ratio between the number of participants and the area of the
experimental zone.

\centerline{}

3 Local density ($\rho$)

\begin{figure}[ht!]

 \centering         
 \includegraphics[scale=0.4]{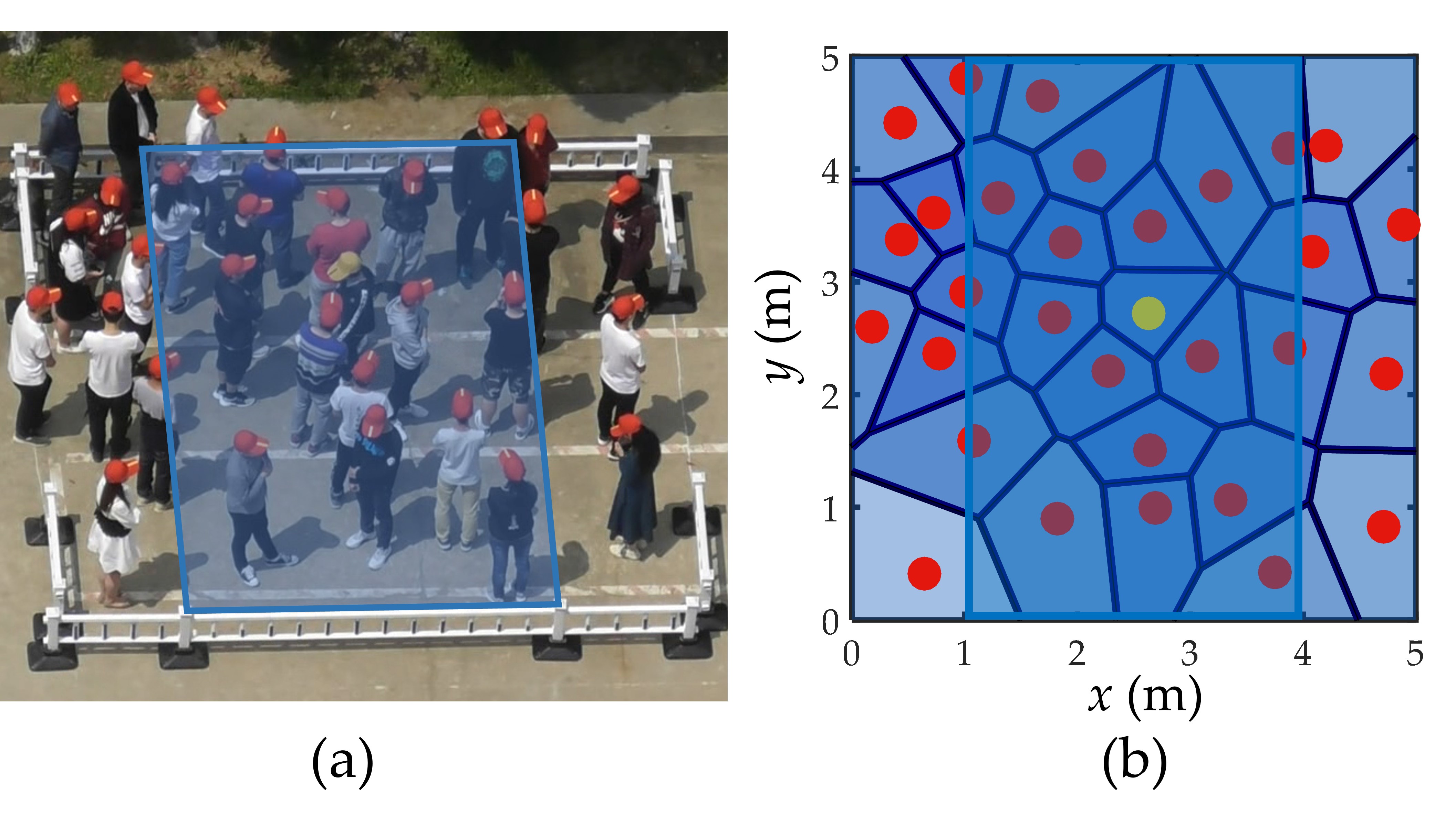}
\caption{Snapshot of the S-L5-N30 experiment and its corresponding Voronoi diagram. (a) The snapshots of the crossing pedestrian in static crowd
conditions, (b) the corresponding Voronoi diagram, where the yellow dot indicates the crossing pedestrian. Correspondingly, the red dots indicate
the stationary volunteers in the experimental area, the color shades of the Thiessen polygons indicate the size of the space occupied, and the blue
rectangular boxed range shows the measurement area (For interpretation of the references to color in this figure legend, the reader is referred to the
web version of this article).
\label{fig4}}
\end{figure}

In contrast to the global density, the local density represents the localized congestion level of pedestrians. The calculation of local
density for crossing pedestrians was performed using the Voronoi diagram method, which has been widely utilized for quantitative
measurements in pedestrian experiments \citep{steffen2010methods} and the expansion of physical models \citep{xiao2016pedestrian}. Fig.\ref{fig4} illustrates
a snapshot of the S-L5-N30 experiment, where red dots indicate the locations of pedestrians within the experimental area, and
yellow dots represent the locations of crossing pedestrians. The Voronoi diagram was generated based on the positions of all pedestrians
within the experimental area at each moment. Subsequently, the area of the Thiessen polygon occupied by pedestrians was
calculated and defined as the local area within the measurement area. The local density of individuals was then determined by Eq.(\ref{2}).

\begin{equation}\label{2}
{{\rho }_{t}}=\frac{1}{{{A}_{t}}}
\end{equation}

where $A_{t}$ signifies the local area occupied by the individual at moment $t$.

\centerline{}

4 Flow rate ($J$)

The flow rate is defined by Eq.(\ref{3}), where it is calculated as the product of the local density of pedestrians and their corresponding
velocities. It quantifies the pedestrian flow characteristics within the localized space.

\begin{equation}\label{3}
{{J}_{t}}={{v}_{t}}\cdot {{\rho }_{t}}
\end{equation}

\centerline{}

5 Eccentricity attentional angle ($\phi$)

To investigate the interference of pedestrians ahead on pedestrian dynamic, we have introduced the concept of the eccentricity attentional angle, denoted as $\phi$ (as shown in Fig.\ref{fig3}(b)). In the absence of eye movement, the eccentricity visual angle within the
Horizontal Field of View (HFV) for humans ranges from 60 to 70 degrees, with the visual field extending beyond 90 degrees and up to
115 degrees on the temporal side \citep{strasburger2020seven}. Considering the head movements of pedestrians during cross, the range of
this eccentricity visual angle is expected to be even larger.

The range of attention, on the other hand, is influenced by various factors
such as visual peripheral distortion, age, velocity, and crowding interference, which can impact the pedestrians’ eccentricity attentional
angle \citep{vonthein2007normal,albonico2018focusing,kewan2022transient}. For scenarios without eye movement (eccentricity
visual angle of 62.5 degrees, HFV of 135 degrees) and scenarios with eye and head movement (eccentricity visual angle of
135 degrees, HFV of 270 degrees), we define the attentional range as the inner 2/3 region closer to the center. In these cases, the
corresponding eccentricity attentional angle is 45 degrees (Horizontal Field of Attention (HFA) of 90 degrees) and 90 degrees (HFA of
180 degrees).

\centerline{}

6 Deviation angle ($\theta$)

The concept of deviation angle was introduced to characterize the angle between the pedestrians’ transient orientation and the
desired orientation, expressed as:
\begin{equation}\label{4}
\theta =\frac{\arccos \left( \frac{{\bm{o}_{des}}\cdot {\bm{o}_{i}}}{\left\| {\bm{o}_{des}} \right\|\cdot \left\| {\bm{o}_{i}} \right\|} \right)\cdot {{180}^{\circ }}}{\pi }
\end{equation}

where, $\bm{o}_{des}$ signifies the desired orientation of the cross pedestrian, and oi indicates the transient orientation of the cross pedestrian $i$.
The desired orientation of pedestrian crossing is determined as the vector pointing from the pedestrian’s current location coordinate
towards the closest exit coordinate.

Based on the location of the upper and lower endpoints of the exit, the experimental area can be
divided into three distinct regions: the upper, middle, and lower regions. When a crossing pedestrian is located in the middle region (as
shown in Fig.\ref{fig3}(c)), their desired orientation corresponds to the vertical direction leading from their position towards the exit.
Conversely, if the pedestrian is situated in the upper region (or lower region, as depicted in Fig.\ref{fig3}(d)), their desired movement
orientation aligns with the vector formed by their position coordinates pointing towards the coordinates of the upper (or lower)
endpoint of the exit.
From a quantitative perspective, even properties that can only be measured by motion sensing could be obtained indirectly by
analyzing trajectories \citep{feliciani2018estimation}. It is essential to note that neglecting pedestrians’ body rotation during avoidance or
detour processes may result in the inaccuracy of raw data in this paper, as this body rotation is considered a crucial form of collision
avoidance movement.
\centerline{}

7 NNRD ($d$)

The NNRD, symbolized as $d$, denotes the Euclidean distance between a crossing pedestrian and its nearest neighboring pedestrian within its HFA. The schematic illustration of this concept can be found in Fig.\ref{fig3}(b), expressed as:

\begin{equation}\label{5}
d=\left\| {\bm{x}_{i}}-\bm{x}_{nei} \right\|
\end{equation}

where $\bm{x}_{i}$ denotes the coordinates of the crossing pedestrian $i$, while $\bm{x}_{nei}$ represents the coordinates of the nearest neighbor pedestrian within the HFA of crossing pedestrian $i$.

\centerline{}

8 NNRA ($\alpha$)

We denote $\alpha$ as the angle between the transient orientation of the crossing pedestrian $i$ and the transient orientation of $i$’s nearest
neighbor pedestrian(within HFA). The schematic diagram of this angle is illustrated in Fig.\ref{fig3}(a). The magnitude of  $\alpha$  represents the
potential conflict. When  $\alpha$  = 0, it indicates that the crossing pedestrian and the nearest neighbor pedestrian(within HFA) have the same
movement orientation, resulting in a lower probability of conflict occurrence. On the other hand, if  $\alpha$  = 180 degrees, it signifies that the
crossing pedestrian and the nearest neighbor pedestrian(within HFA) have opposite movement orientations, implying that the
pedestrian may need to take evasive actions to avoid potential conflicts. The physical interpretation of this angle is provided by Eq.(\ref{6}).

\begin{equation}\label{6}
\alpha =\frac{\arccos \left( \frac{{\bm{o}_{nei}}\cdot {\bm{o}_{i}}}{\left\| {\bm{o}_{nei}} \right\|\cdot \left\| {\bm{o}_{i}} \right\|} \right)\cdot {{180}^{\circ }}}{\pi }
\end{equation}

where $\bm{o}_{nei}$ represents the transient orientation of the nearest neighbor pedestrian within the HFA of crossing pedestrian $i$.

\centerline{}

9 Angular velocity ($\omega$)

The angular velocity represents the rate of angular change of the crossing pedestrian within a sampling time gap $\tau$, and the angular velocity $\omega$ of the crossing pedestrian $i$ at moment $t$ is defined as:

\begin{equation}\label{7}
\omega =\frac{\arccos \left( \frac{{\bm{o}_{i,t+\tau /2}}\cdot {\bm{o}_{i,t-\tau /2}}}{\left\| {\bm{o}_{i,t+\tau /2}} \right\|\cdot \left\| {\bm{o}_{i,t-\tau /2}} \right\|} \right)\cdot {{180}^{\circ }}}{\pi \cdot \tau }
\end{equation}

We will employ a consistent variable notation in the subsequent analysis based on the defined observable variables presented above.

\section{Does dynamism of crowd persist?} \label{section5}

Why does the collective dynamism of crowds persist? Before addressing this question, it is imperative to ascertain the questions’
ontological status. Hence, the present section empirically examines the existence of persistent dynamism in crowds.

\subsection{Smokescreen of measurements} \label{subsection5.1}
Before analyzing and discussing the empirical results, explaining the differences in methodologies of pedestrian measurement is
essential. This clarification is necessary to avoid erroneous interpretations of the raw data. 

The early studies of pedestrian dynamics
focused on fundamental diagrams, a paradigm that originated from traffic research. Pedestrians, especially those in single-file arrangements,
exhibit movement characteristics similar to vehicles, showing similar trends in fundamental diagrams. As pedestrian
research delved deeper, a strong correlation between velocity and space ahead (headway) rather than the occupied circular space
(local density) has been discovered.
The relationship between pedestrian velocity and the corresponding headway has been the subject of numerous experimental
investigations \citep{cao2016pedestrian,wang2021pedestrian}. 

Generally, a critical headway value is observed below which pedestrian movement
ceases. This threshold also corresponds to the point at which pedestrian velocity reaches nullity in single-file experiments. The critical
headway value is typically discerned via linear regression of headway data linked to non-free-flow velocities, wherein the intercept of
the regression line signifies the critical headway. The linear relationship between instantaneous velocity and headway has been
revealed in numerous studies, which is consistent with empirical perception, where the velocity is proportional to the distance ahead
with a fixed coefficient $S$. And, in most empirical statistics, the relationship between velocity and local density is nonlinear.

\begin{figure}[ht!]
\plotone{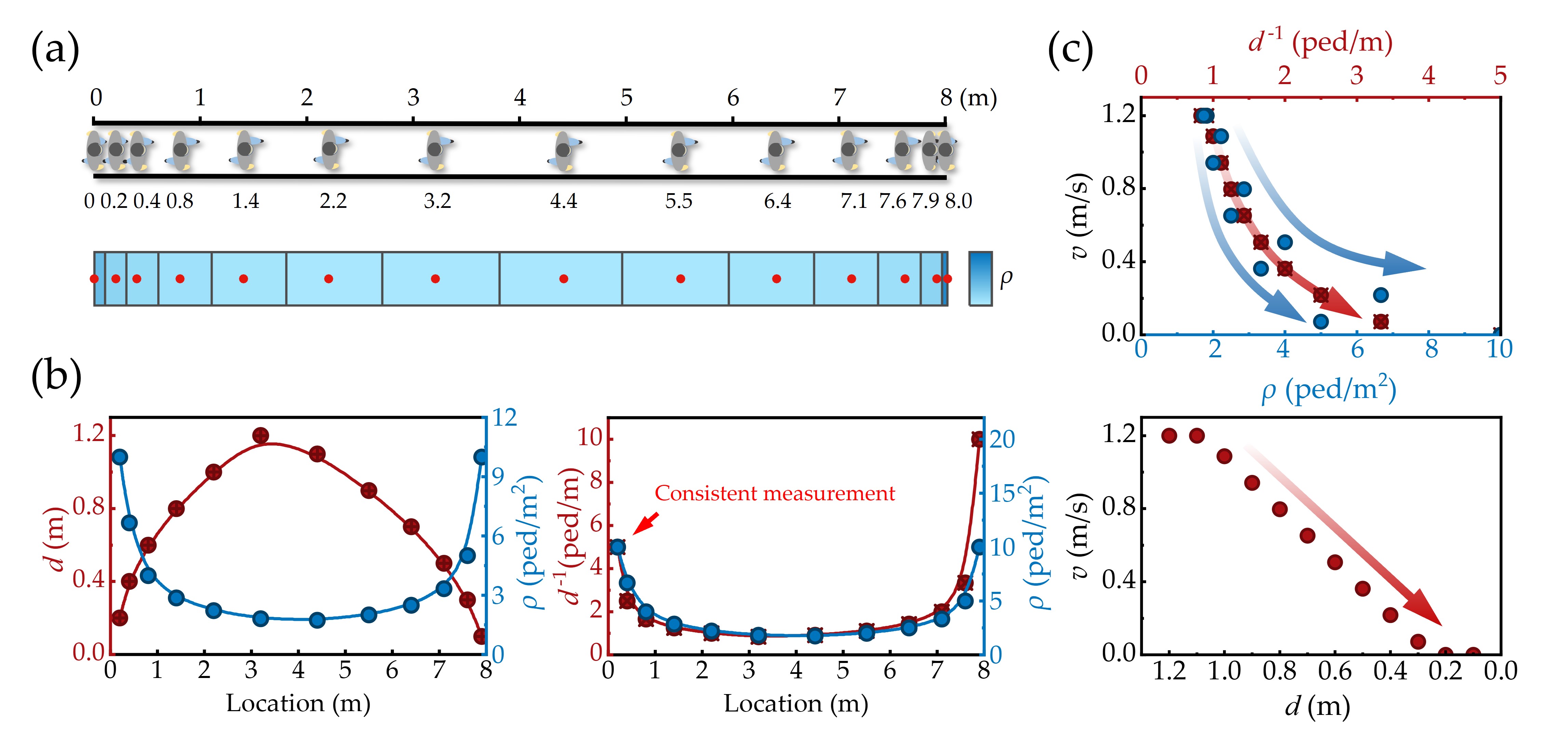}
\caption{(a) Snapshots of pedestrian experiments and corresponding representations of local density. The measurement of local density also adopts
the Voronoi diagram method. (b) Left: Distribution variations of different positions of pedestrians, $d$, and $\rho$. Right: Distribution variations of different
positions of pedestrians, $d^{-1}$, and $\rho$. There exists a conversion ratio of 0.5 (corridor width) between $d^{-1}$ and $\rho$, enabling the comparison of the two
measurement methods in different scales of coordinates.
\label{fig5}}
\end{figure}

A simple scenario in Fig.\ref{fig5} has been presented to understand this variation, illustrating a snapshot of a single-file experiment with a
corridor of 0.5 m × 8 m. The coordinates of each pedestrian within the channel are provided in the figure. In crowd experiments,
precise pedestrian headway data is unattainable. Thus, statistical analysis of the NNRD affords an approximation of the relationship
between pedestrian crossing velocity and space ahead. In the experiment of single-file context, NNRD and headway are deemed
formally equivalent (headway is defined as the distance between the pedestrian and his predecessor). Here, we postulate a linear
relationship between pedestrian velocity and NNRD, with the corresponding expression provided by \citet{cao2016pedestrian}:

\begin{equation}\label{8}
v=\frac{d-I}{S},(0<v<{{v}_{\max }})
\end{equation}

where,$S=\frac{\delta d}{\delta v}=0.69$ serves as the proportionality constant, and $I$ = 0.25 represents the upper limit of forward distance for maintaining a velocity of 0, and $v_{\max}$ = 1.2 (m/s) denotes the maximum speed.
Pedestrian data at both ends of the corridor are excluded to mitigate the influence of boundary conditions. 

This allows us to obtain a
velocity sequence demonstrating a linear relationship with NNRD, as depicted in Fig.\ref{fig5}(c). Fig.\ref{fig5}(b) displays the observable distribution
of NNRD($d$), the reciprocal of NNRD($d^{-1}$), and local density ($\rho$) at each positions. The arrow points to pedestrian data with a
positional coordinate of 0.2 m, where the forward and backward distances are equal. Therefore, the measurement of $d^{-1}$ and $\rho$ is
consistent, without error. Point this as a reference, we observe that the local density of pedestrians on the left side of the passage is
overestimated (their NNRD is greater than that of pedestrians behind them), while the local density of pedestrians on the right side is
underestimated (their NNRD is smaller than that of pedestrians behind them), as shown in Fig\ref{fig5} (b) on the right. This discrepancy leads
to the domain in the distribution of the velocity-density relationship, where the upper boundary represents the overestimated density
data, and the lower boundary indicates the underestimated density data, with velocity and density following an inverse proportional
function, as presented in Fig.\ref{fig5}(c).

Thus, it’s challenging to conclude that pedestrians persist dynamism in high-density conditions through velocity-density diagram.
Illusion caused by differentiated measurements makes it easily observed that pedestrians persist dynamism in high-density conditions.
However, this observation is the result of inaccurate evaluations because we have already defined the maximum NNRD for pedestrian
standstill in the empirical formula ($I$ = 0.25). Therefore, comparatively speaking, the relationship between velocity and NNRD better captures the nature of pedestrian movement patterns.

In the following, the results presented by the two measurement methods will be discussed to demonstrate the proposition’s validity regarding the crowd’s persistent dynamism.

\subsection{Density-velocity relation} \label{subsection5.2}

\begin{figure}[ht!]
\plotone{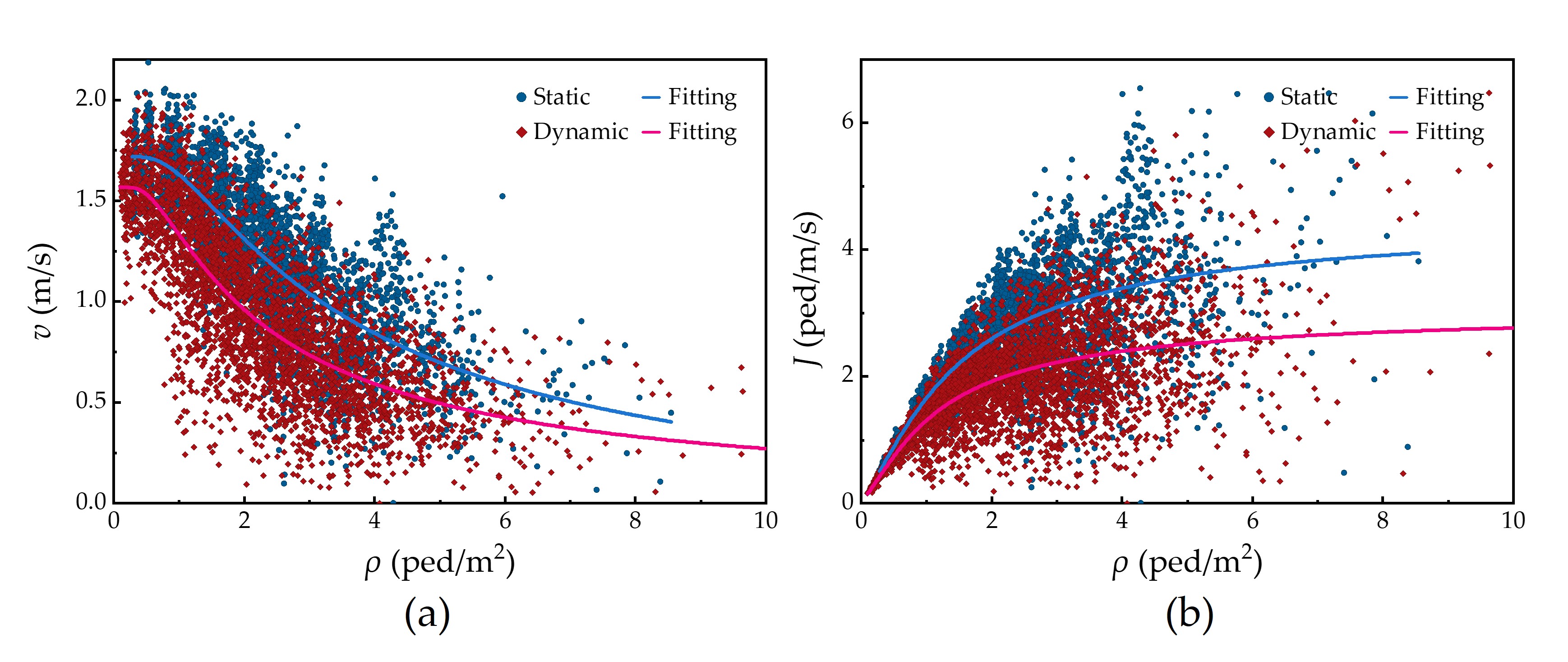}
\caption{Fundamental diagrams. The variation of speed and flow rate was fitted using the Weidmann equation, and the results are given in Table \ref{table4}.
\label{fig6}}
\end{figure}

Based on the conducted experiments described above, we can analyze the movement data of individuals crossing in different crowd
patterns. Firstly, we examined the relationships of fundamental diagrams to investigate the difference displayed by pedestrians during
cross. These variations are illustrated in Fig.\ref{fig6}. Fig.\ref{fig6}(a) presents the velocity-density plot of the fundamental diagram for the two
tested patterns. A decrease in velocity is observed as density increases, and each data set aligns well with the Weidmann equation
\citep{weidmann1993transporttechnik}, which is expressed as:

\begin{equation}\label{9}
v={{v}_{\max }}\left\{ 1-\exp \left[ -\gamma \left( \frac{1}{\rho }-\frac{1}{{{\rho }_{\max }}} \right) \right] \right\}
\end{equation}

where $v_{\max}$ denotes the maximum speed (or the free walk speed) and ${\rho }_{\max }$ refers to the maximum density where the speed drops to zero.
Fig.\ref{fig6}(b) illustrates the flow rate versus density representation in the fundamental diagram. The flow equation can be derived by
multiplying Weidmann’s equation with the density \citep{feliciani2018measurement,cristiani2014multiscale}:

\begin{equation}\label{10}
J=\rho \cdot {{v}_{\max }}\left\{ 1-\exp \left[ -\gamma \left( \frac{1}{\rho }-\frac{1}{{{\rho }_{\max }}} \right) \right] \right\}
\end{equation}

Fig.\ref{fig6} and Table \ref{table4} illustrate the trends observed in the variation of the fundamental diagrams. Compared with dynamic context, cross pedestrians exhibit higher free-flow speeds within static context. Interestingly, even with the same local density, pedestrians can achieve higher speeds and flow rates when crossing static crowds. These findings reveal that relying solely on local density may not
provide sufficient accuracy for classifying Pedestrian Level of Service (PLOS) \citep{jia2022revisiting}.
In some cases of high local density, it is evident that pedestrians can still maintain movement. This observation aligns with the findings of Helbing \citep{helbing2007dynamics,yu2007modeling}. Based on the presented results, appears that the argument of dynamism of the crowd can be established. However, it was noted that the local density measurement distorts the description of movement space. Therefore, the existing evidence yet insufficient to support this argument.

\begin{table}[!htb]
\centering
\caption{Coefficients result from fitting the Weidmann equation to the results of the experiment. Speed and flow rate are fitted separately, resulting in different coefficients. Data in this table refer to the graph in Fig.\ref{fig6}.}
\resizebox{0.6\columnwidth}{!}{
\begin{tabular}{llllll}
\toprule
Function                   & Experiment & $v_{\max}$ & $\rho_{\max}$       & $\gamma$ & $R^{2}$ \\
\midrule
\multirow{2}{*}{Speed}     & Static     & 1.718      & 34.304              & 3.055    & 0.636   \\
                           & Dynamic    & 1.567      & $3.564 \times 10^{20}$ & 1.900    & 0.616   \\
\multirow{2}{*}{Flow rate} & Static     & 1.798      & 707.129             & 2.568    & 0.560   \\
                           & Dynamic    & 1.511      & $3.279 \times 10^{21}$ & 2.023    & 0.431   \\
\bottomrule
\end{tabular}
}
\label{table4}
\end{table}

\subsection{Distance-velocity relation} \label{subsection5.3}

\begin{figure}[ht!]
\plotone{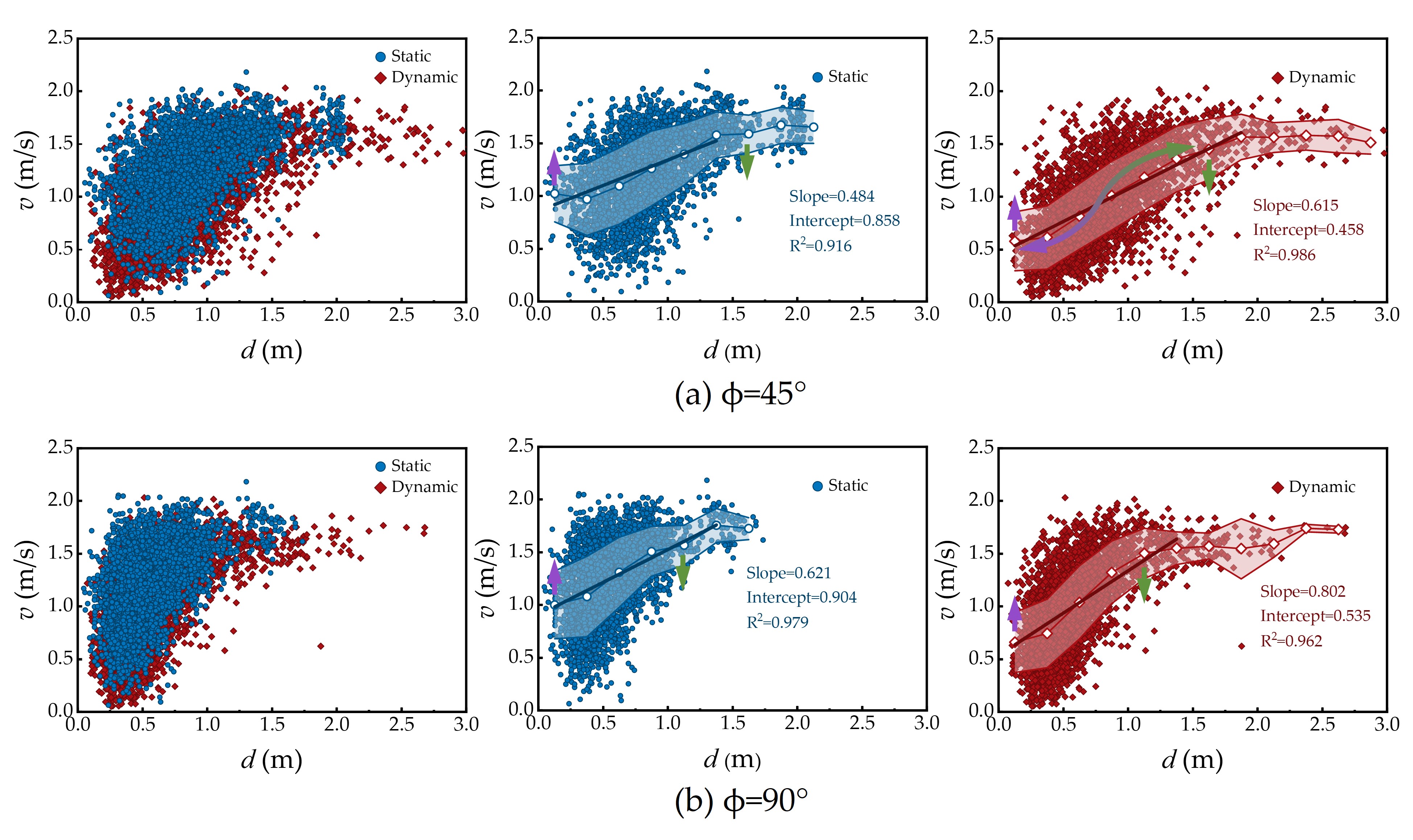}
\caption{Relationship between the instantaneous velocity of cross pedestrians versus the NNRD is examined under different Eccentricity attentional
angle conditions, (a)$\phi$= 45 degrees, (b)$\phi$= 90 degrees.
\label{fig7}}
\end{figure}

The scatter plot of instantaneous velocity versus NNRD, portrayed in Fig.\ref{fig7}, delineates the instantaneous velocity variations with
the NNRD under varying eccentricity attentional angles ($\phi$ = 45 degrees and $\phi$ = 90 degrees). The velocity fluctuations of cross pedestrians
can be partitioned into two stages: an initial linear-like augmentation stage during non-free-flow conditions and a subsequent
stable free-flow stage.

Fig.\ref{fig7} illustrates a linear-like augmentation stage, but the inconstant velocity change can be observed upon comparing the mean
data and the distribution of the fitted line. At low NNRD domain and speeds approaching free flow velocity, the velocity change
experiences a decline. Within the velocity domain approaches free flow velocity, this deceleration has been previously identified in
relevant studies \citep{jelic2012fundamental}, and may attributed to individual differences. As participants’ free velocities differ, with an increase
in NNRD, those with lower free velocities reach their free velocities earlier, thereby reducing the overall speed gain.

Within the low NNRD domain, why does the deceleration of velocity change? A reasonable speculation is that some pedestrians opt for detours
under spatial constraints. Cross pedestrians exhibit flexible navigation within the crowd, facilitating unconstrained detours and
overtaking maneuvers. This absence of rigid spatial constraints diverges to a certain degree from the outcomes of single-file experiments where overtaking and lateral movements are generally precluded. As the NNRD of cross pedestrians approximates zero,
the pedestrian movement continues at an approximate minimum speed denoted by the intercept. This indicates that pedestrians can
maintain movement even in extreme spatial constraints ($d$ → 0), providing empirical evidence to support the notion of dynamism in crowds.

\section{Micro-analysis} \label{section6}
Does the dynamism of the crowd persist? In the previous section, empirical evidence was presented. In this section, our research
revisits the core question: Why does the collective dynamism of the crowd persist? Empirical speculation points to the notion of
pedestrians’ detour behavior, which currently lacks evidence. In this section, we observe the crossing behaviors of pedestrians within
crowds. Clearly, pedestrians exhibit distinct movement characteristics in static and dynamic context and under low or high spatial
constraints. By unraveling the intricacies of pedestrian dynamics, these studies are poised to contribute significantly to our comprehension
of how pedestrians adapt their behaviors based on crowd patterns. These qualitative characteristics will provide us with new
perspectives, ultimately guiding us toward answers.

\subsection{Crossing trajectory}\label{subsection6.1}

Pedestrian trajectories represent the most discernible and informative features of movement. In Fig.\ref{fig8}, we present illustrative
diagrams depicting the trajectories of pedestrians across diverse conditions. These diagrams offer visual representations that enable a
comprehensive understanding of the complex and dynamic movement characteristics during the crossing process.

\begin{figure}[ht!]
\plotone{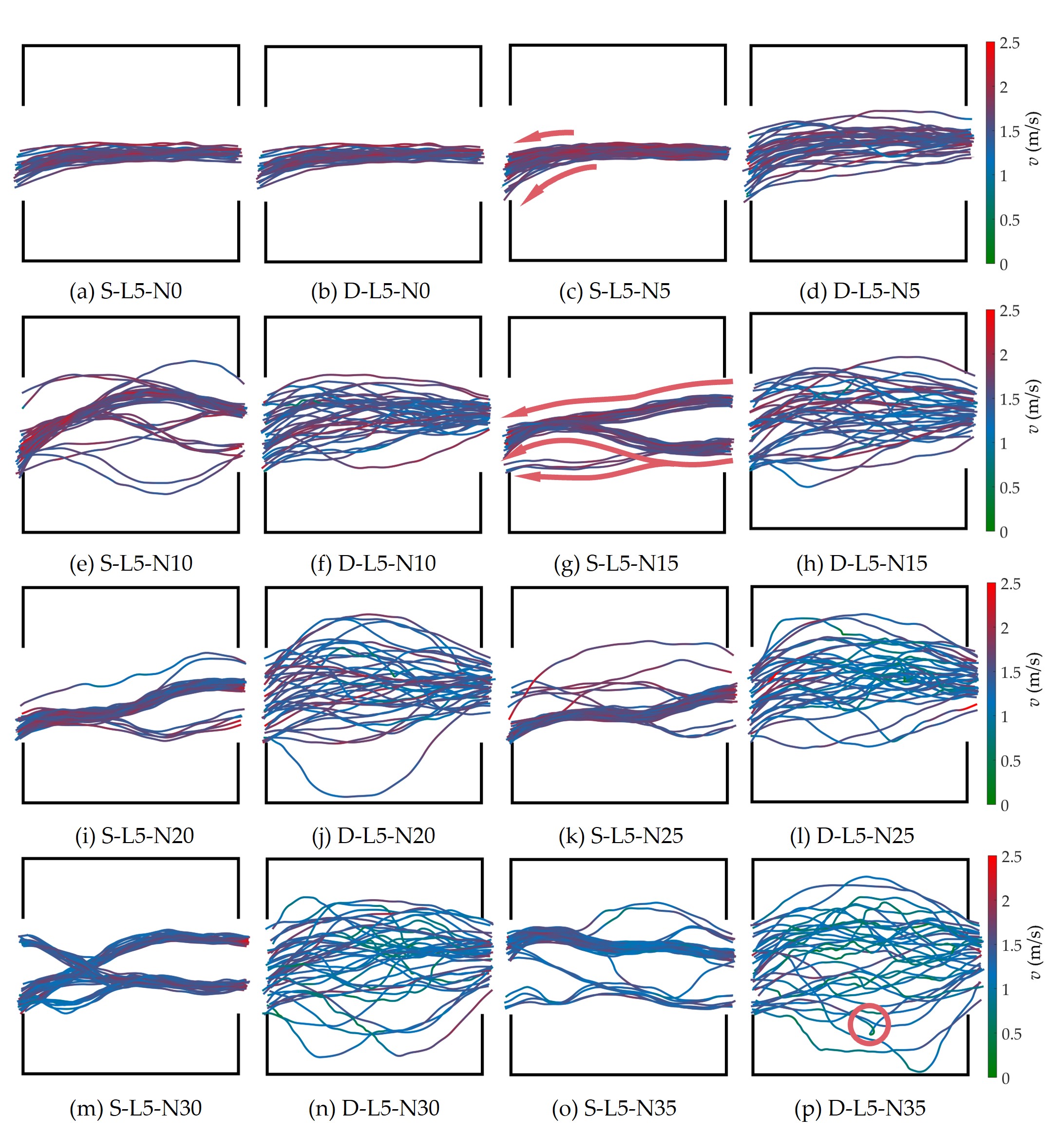}
\caption{Pedestrian crossing trajectory diagrams.
\label{fig8}}
\end{figure}

From Fig.\ref{fig8}, we observed that when pedestrians cross through static crowds, their trajectories exhibit higher concentration, even in
scenarios with low global density. Notably, as pedestrians approach the exit of the experimental area, there is a clear downward trend
in their trajectory as they move toward the holding area (indicated by the short arrow in Fig.\ref{fig8}(c)). Additionally, the trajectory diagrams
reveal certain self-organized behaviors. Pedestrians spontaneously form cross-channels to avoid stationary pedestrians (represented by the long arrow in Fig.\ref{fig8}(g)).

Conversely, when pedestrians cross dynamic crowds, distinct features such as backward
detours and spiral turning behaviour emerge (as observed in the circles in Fig.\ref{fig8}(p), Fig.\ref{fig8c}(r), and Fig.\ref{fig8c}(v)), whereas these phenomena
are absent in static crowd experiments. A common feature observed during the crossing process is the noticeable variation in speed
induced by changes in turn or detour maneuvers (highlighted by the rectangle in Fig.\ref{fig8c}(aa) and Fig.\ref{fig8c}(cc)). In the subsequent section,
we delve into the microscopic analysis of trajectory, focusing on cross-channel formation, spatial characteristics, backward detours,
and spiral-shaped turns.

\begin{figure}[ht!]\ContinuedFloat
\plotone{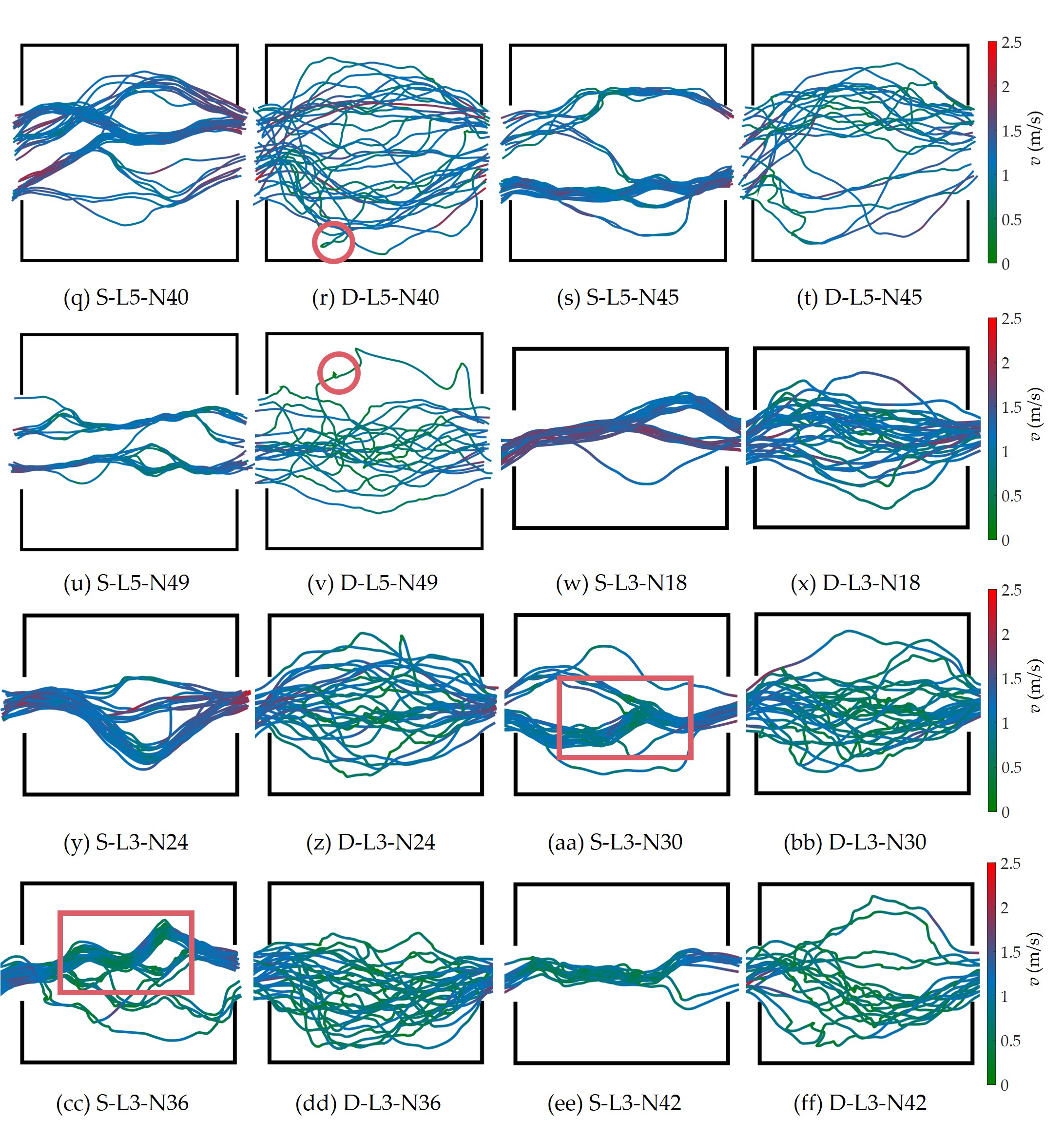}
\caption{Pedestrian crossing trajectory diagrams (continue).
\label{fig8c}}
\end{figure}

\subsection{Cross-channel formation mechanism}\label{subsection6.2}

\begin{figure}[ht!]
\plotone{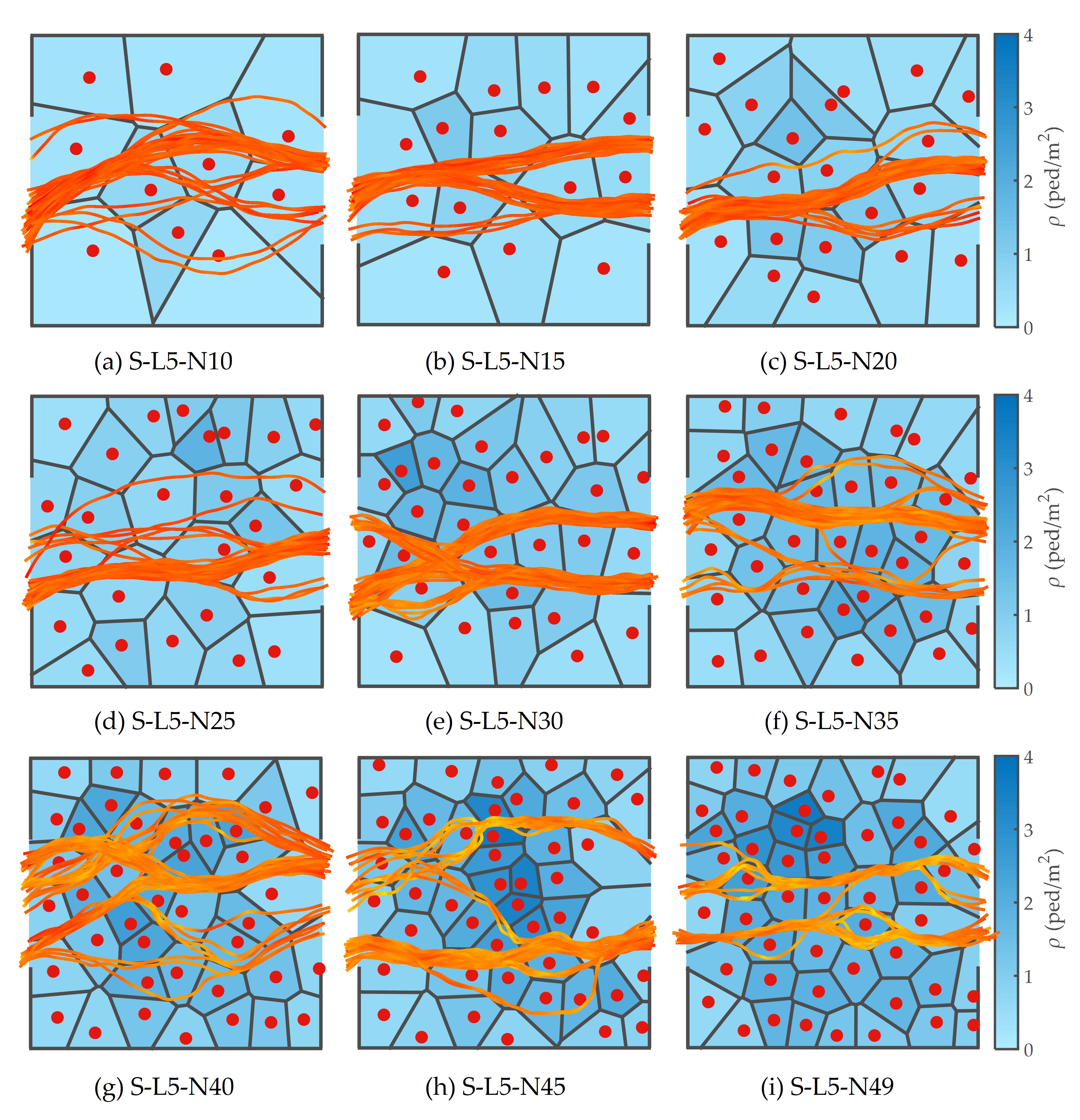}
\caption{The Voronoi diagrams in different experiments are superimposed on the cross trajectories, the red dots indicate stationary pedestrians in the
experimental area, and the data interpolation method is applied to the filling of the local area fields. 
\label{fig9}}
\end{figure}

In this part, we delve into a thorough investigation of pedestrian crossing strategies, aiming to unravel the underlying mechanisms
governing the formation of cross channels. Fig.\ref{fig9} illustrates the Voronoi diagram’s distribution within the experimental area, spanning
from experiment S-L5-N10 to experiment S-L5-N49. The yellow curve denotes the trajectory of each pedestrian crossing. The interpretations
drawn from Fig.\ref{fig9}(a) through Fig.\ref{fig9}(d) indicate that when the global density is below 1 $ped/m^{2}$ within the specified
experimental domain, the spatial expanse utilized by the individuals maintains a notably high level. Under such conditions, pedestrians
aspire to cross space by adhering to the shortest and most direct path, leading to a decrease in instances of detours and deviations
along their projected trajectories.
As the global density within the experimental domain escalates, the localized space accessible to individuals diminishes, obliging
crossing pedestrians to execute recurrent turns to strategize their transit routes, as depicted in Fig.\ref{fig9}(e)-(i). Remarkably, as showcased
in Fig.\ref{fig9}(g), (h), and (i), pedestrians exhibit evasion of localized high-density zones. Anticipation behaviour of pedestrians facilitates
their prompt evasion of these high-density zones along their trajectories, suggesting the necessity to forfeit immediate choices to
realize an optimal route. This mechanism is considered instrumental in the genesis of lanes within bi-directional pedestrian flows \citep{murakami2021mutual}. As depicted in Fig.\ref{fig9}(h), pedestrians opt for larger-scale detouring instead of penetrating through the lower-density area at the entrance. The purpose of this maneuver is to avoid the "human wall" at the central section of the experimental area.

In static context, a tendency towards analogous movement strategies is displayed by pedestrians, with a priority on circumventing
high-density zones. Therefore, can it be inferred from this that the lower spatial constraints drive the formation of crossing channels in
static crowds? Existing results lack quantitative support, and statistical regularities will provide evidence subsequently.

\centerline{}
\subsection{Micro-analysis in detour mechanism}\label{subsection6.3}

\begin{figure}[ht!]
 \centering         
 \includegraphics[scale=0.56]{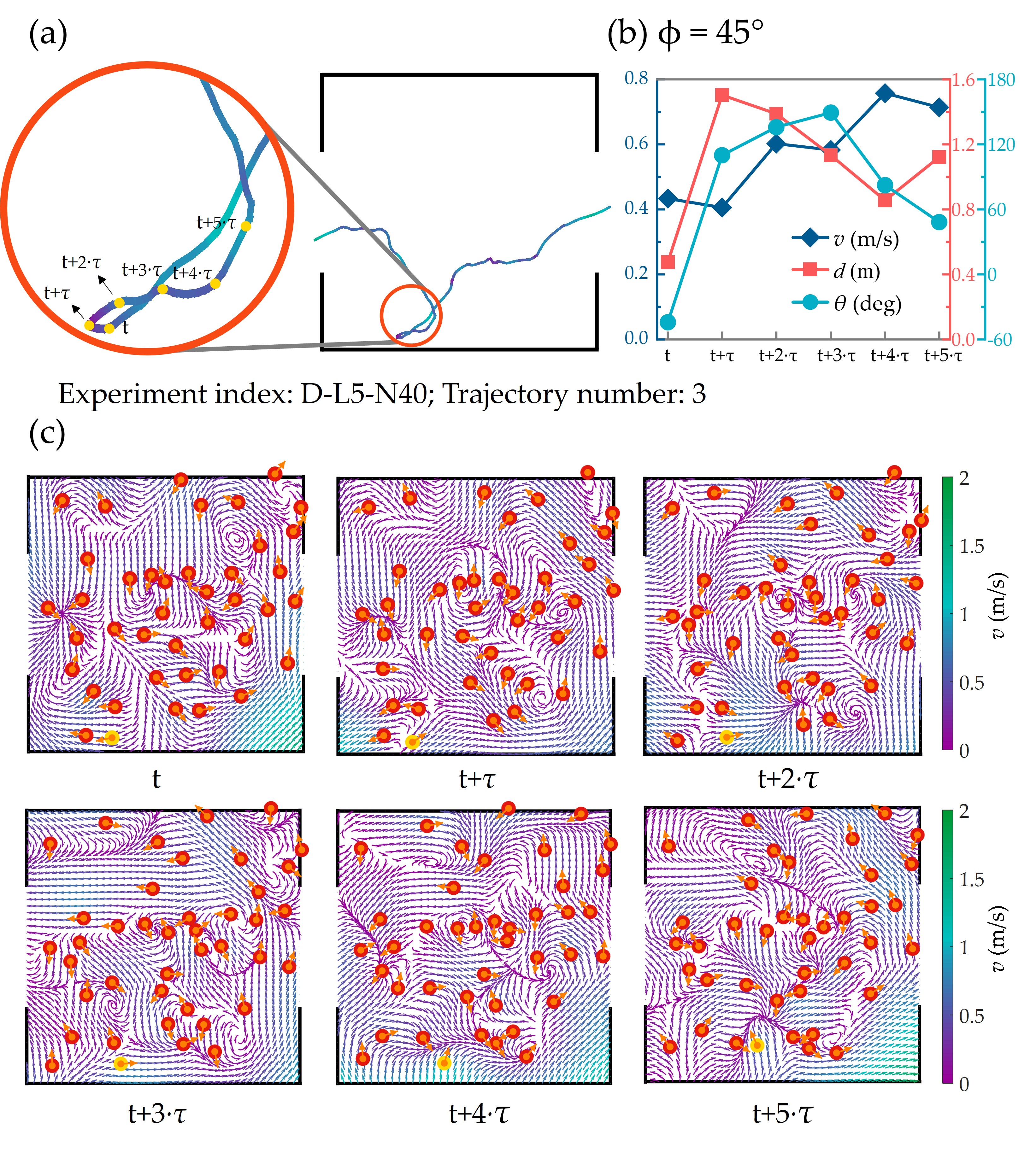}
\caption{(a) The experiment M-L5-N40 exhibits a discernible backward detour behavior, with the third trajectory indicating the path taken by the
cross pedestrian. (b) The corresponding fluctuations in instantaneous velocity, deviation angle, and NNRD of cross pedestrians at each discrete
sampling moment are graphically illustrated. It’s worth mentioning that discrepancies in velocity values at corresponding sampling moments in (a)
and (b) are attributable to variations in the sampling time interval. In the trajectory graphs, velocity sampling is executed for each frame at a 0.04 s
time interval. (c) The spatial distribution of the pedestrian velocity field at distinct moments is depicted, with red dots symbolizing the position
coordinates of pedestrians within the experimental zone and yellow dots signifying the position coordinates of cross pedestrians. A method of data
interpolation is employed to populate the velocity field. In the upper right corner of each subplot, one pedestrian’s coordinates exceed the boundary
of the experimental area, a projection error ascribed to the varying heights of pedestrians, which has been discussed previously. (For interpretation
of the references to color in this figure legend, the reader is referred to the web version of this article.)
\label{fig10}}
\end{figure}

\begin{figure}[ht!]
 \centering         
 \includegraphics[scale=0.56]{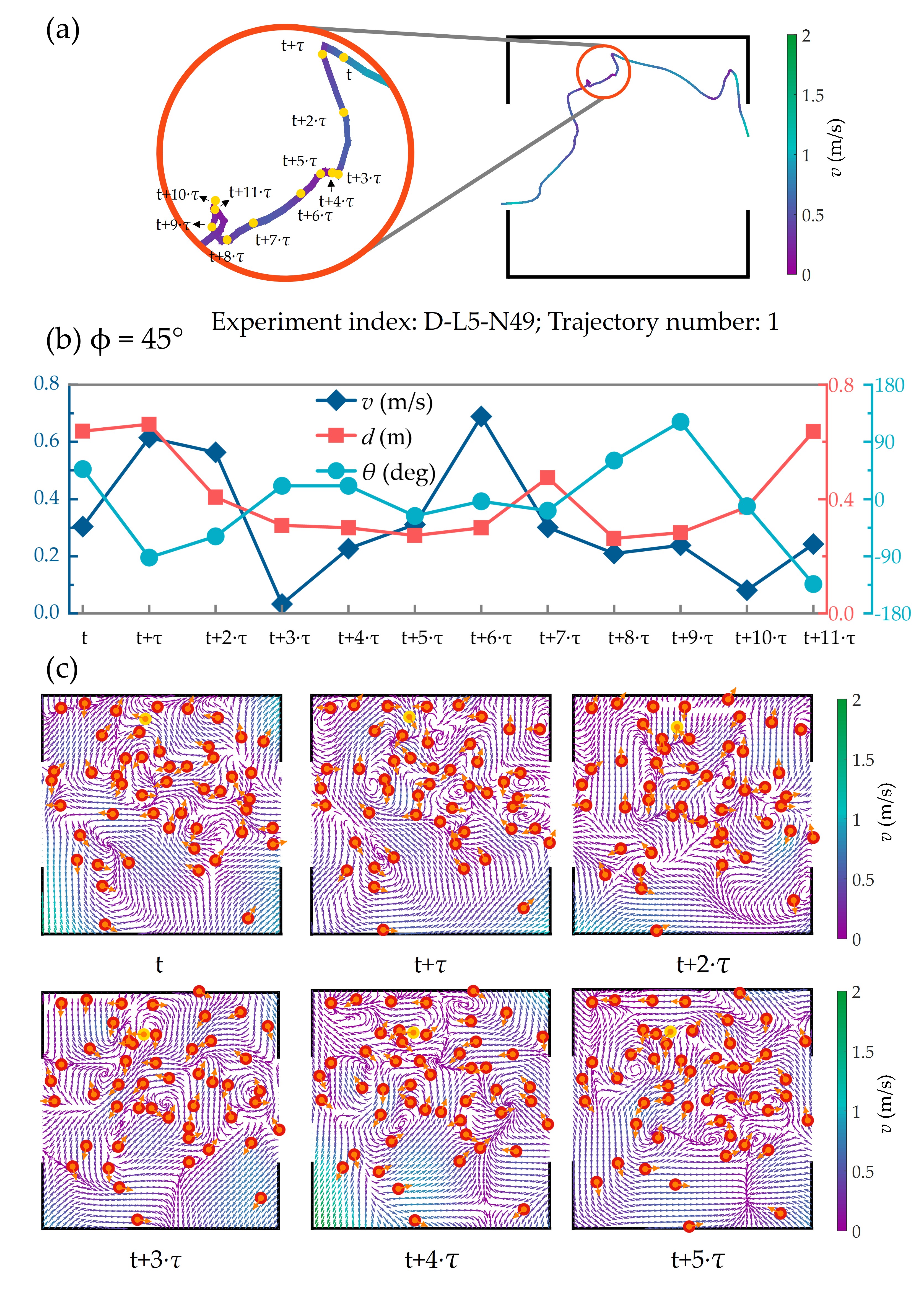}
\caption{(continued).
\label{fig11}}
\end{figure}

\begin{figure}[ht!]\ContinuedFloat
 \centering         
 \includegraphics[scale=0.56]{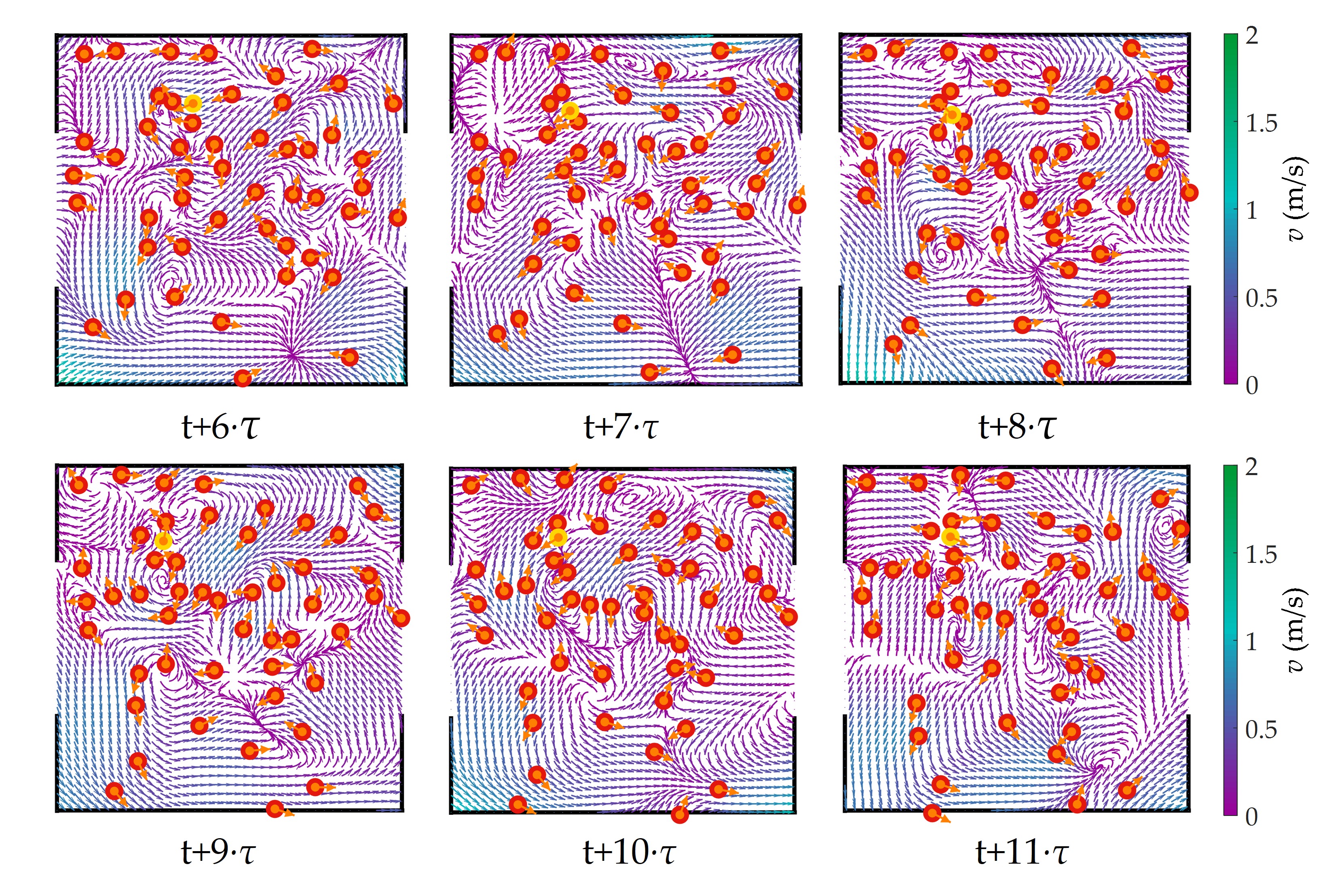}
\caption{(a) The observed backward detour behavior in the M-L5-N49 experiment, indicated by trajectory number 1, signifies the first trajectory
of the cross pedestrian. (b) The corresponding changes in instantaneous velocity, deviation angle, and NNRD of cross pedestrians at each sampling
moment are graphically represented. (c) The spatial distribution of the pedestrian velocity field at different moments is depicted. Notably, in the
upper right corner of each subplot, there is a pedestrian whose coordinates extend beyond the boundary of the experimental area, a projection error
ascribed to the height disparities among pedestrians, which has been addressed previously. Fig.\ref{fig11}. (a) The observed backward detour behavior in
the M-L5-N49 experiment, indicated by trajectory number 1, signifies the first trajectory of the cross pedestrian. (b) The corresponding changes in
instantaneous velocity, deviation angle, and NNRD of cross pedestrians at each sampling moment are graphically represented. (c) The spatial
distribution of the pedestrian velocity field at different moments is depicted. Notably, in the upper right corner of each subplot, there is a pedestrian
whose coordinates extend beyond the boundary of the experimental area, a projection error ascribed to the height disparities among pedestrians,
which has been addressed previously. (continued).
\label{fig11c}}
\end{figure}

The above analysis unveiled the emergence of backward detours and spiral-shaped trajectories when pedestrians cross through
dynamic crowds, a phenomenon not observed in experiments involving static crowds. This suggests that these manifestations are likely
precipitated by the velocity field’s influence on pedestrian movement direction. We investigate the velocity field within the experimental
area to analyze the formation mechanism of pedestrian detours, focusing on the process of backward detours.

The intriguing occurrence of cross pedestrians executing backward detours in experiment D-L5-N40 is the initial focus of our
analysis, as depicted by the pink circles in Fig.\ref{fig8c}(r). Fig.\ref{fig10}(a) offers a comprehensive view of the trajectory along with the associated
sampled point data. To scrutinize pedestrian dynamic, we extract the velocity field from the experimental area at six discrete moments,
spanning a total duration of 2.4 s. Fig.\ref{fig10}(b) demonstrates the fluctuations in instantaneous velocity, deviation angle, and NNRD of pedestrians at each moment. Fig.\ref{fig10}(c) visualizes the velocity field at each moment.

Fig.\ref{fig10}(c: $t$) reveals that at moment t, the left side of the cross pedestrian is the boundary, and three pedestrians are positioned rightfront
towards the cross pedestrian, blocking the movement towards the exit. As a result, in the subsequent moment (Fig.\ref{fig10}(c:$t+\tau$)), the
pedestrian reverses orientation to evade the pedestrians ahead. During this detour towards the right rear, the pedestrian encounters an
abrupt change in direction from the pedestrian to its right, creating a conflict with the pedestrian’s current walking trajectory. This
compels the pedestrian to retreat backward, as illustrated in Fig.\ref{fig10}(c:$t+2\tau$). Analysis reveals that cross pedestrians resort to backtracking and circumventing during the cross process due to obstructions ahead. This behavior expands the scope of the backtracking
phenomenon and engenders a spoon-shaped trajectory during the crossing, as exemplified in Fig.\ref{fig10}.

In experiment D-L5-N49, the pedestrians’ spiral turning has been observed, as depicted by the pink circles in Fig.\ref{fig8c}(v). To enable an
exhaustive analysis, we select an extended time interval comprising 12 discrete moments, corresponding to a total duration of 4.8 s.
The fluctuations in instantaneous velocity, deviation angle, and NNRD of pedestrians at each moment are depicted in Fig.\ref{fig11} (b). The
velocity field at each of these moments is illustrated in Fig.\ref{fig11}(c).

At moment $t$, as indicated in Fig.\ref{fig11}(c:$t$), the cross pedestrian movement towards the boundary of the experimental area.
Continuing this orientation would lead to colliding with the boundary. Therefore, the pedestrian initiates a reversal in direction,
turning left for rearward movement in the subsequent moment, as demonstrated in Fig.\ref{fig11}(c:$t+\tau$). After crossing a short distance,
turbulence in the velocity field materializes at the pedestrian’s location, causing a decrease in its velocity, as illustrated in Fig.\ref{fig11}(c:$t+3\tau$) and Fig.\ref{fig11}(c:$t+4\tau$). At the moment $t+5\tau$, the pedestrian alters its orientation and moves towards the exit but reencounters
turbulence at the moment $t+8\tau$ and makes physical contact with a movement pedestrian. Because leading and trailing pedestrians
move in opposing orientations, the cross pedestrian undergoes a spiral turning due to external influence.

The described process exemplifies the complex dynamic during the crossing process, where dense crowds may encounter collisions,
positional conflicts, and physical contact, all influencing the behavior of cross pedestrians collectively. Through scrutinizing the velocity
field within the experimental area, we deduce that pedestrians resort to backward detours to evade impending collisions or
positional conflicts when no feasible forward walking route is available. Consequently, they maneuver through the situation by
backtracking and circumventing the obstacle. The emergence of spiral turning is instigated by turbulence. When the local density is
high and the pedestrians’ orientations are inconsistent, walkers deviate from their intended trajectory due to entanglement within the
crowd, resulting in spiral-shaped trajectories.

\section{Statistic analysis} \label{section7}

In the preceding analysis, we conducted a microscopic examination of pedestrian crossing movements, exploring the formation of
crossing channels and the occurrence of detours. In the subsequent section, we will employ a statistical analysis of pedestrian
movement to elucidate the macroscopic patterns of pedestrian crossings, and the existing questions will be addressed.

\subsection{Spatial analysis}\label{subsection7.1}

\begin{figure}[ht!]
\plotone{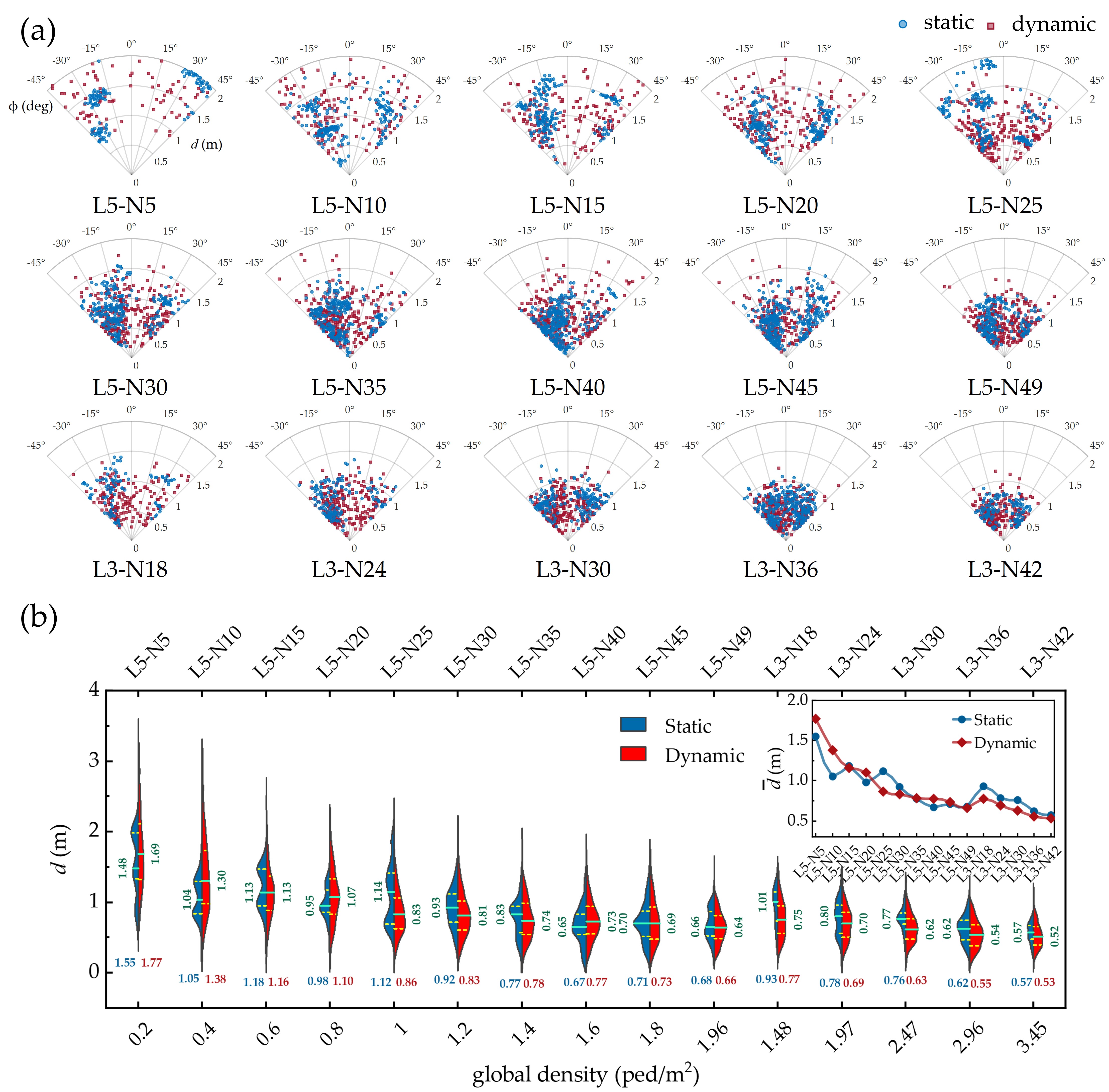}
\caption{$\phi$=45 degrees, (a) Spatial distribution of nearest neighbor when pedestrians were crossing different crowd patterns. (b) The distribution of NNRD under different global densities.
\label{fig12}}
\end{figure}

\begin{figure}[ht!]
\plotone{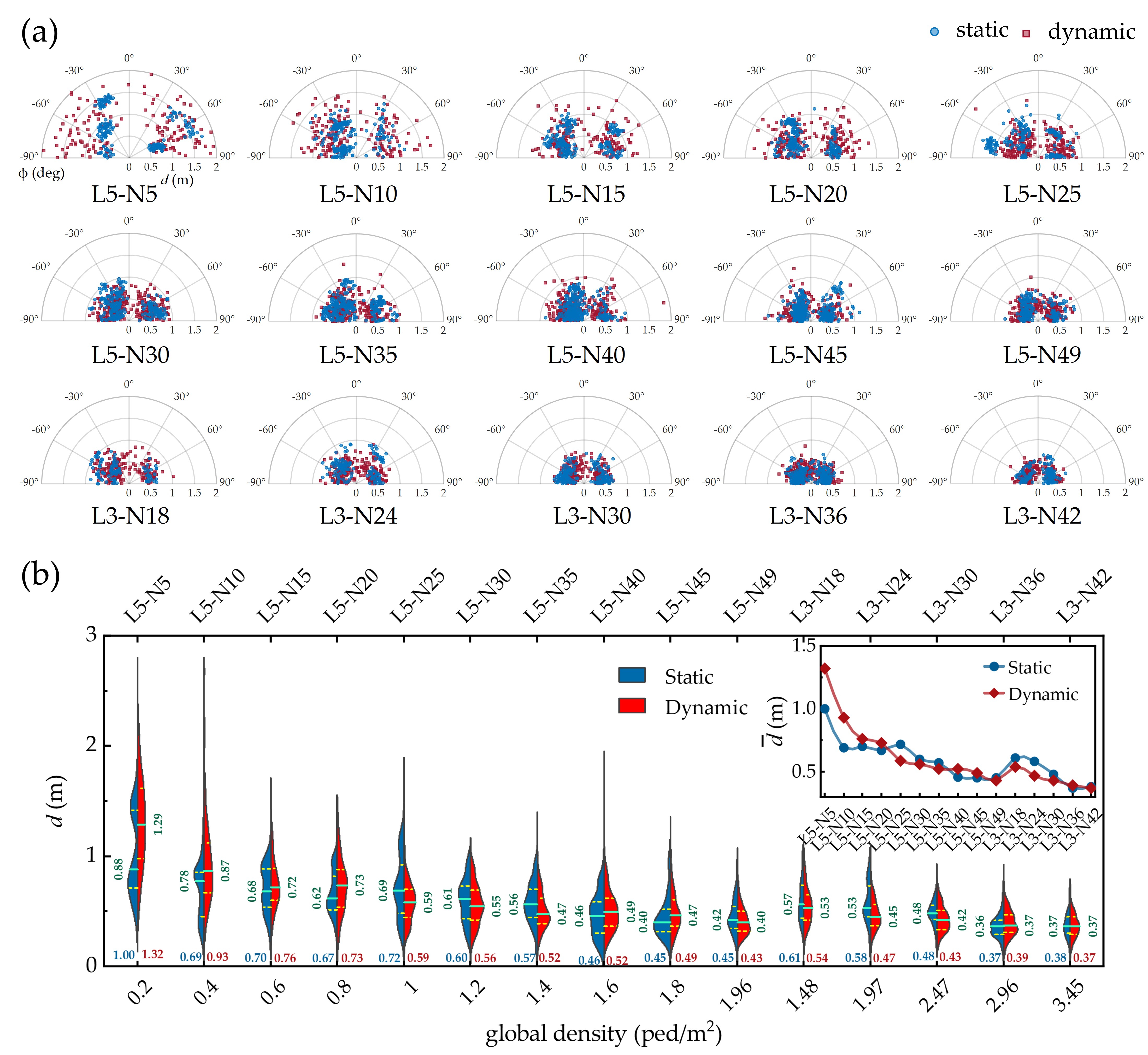}
\caption{$\phi$=90 degrees, (a) Spatial distribution of nearest neighbor when pedestrians were crossing different crowd patterns. (b) The distribution of NNRD under different global densities.
\label{fig13}}
\end{figure}

Statistical analyses were conducted to elucidate the spatial relationship between cross pedestrians and their nearest neighbor.
Fig.\ref{fig12} and Fig.\ref{fig13} delineate the spatial distribution of the nearest neighbor of cross pedestrians within the HFA for eccentricity
attentional angles of 45 degrees and 90 degrees, respectively.

Fig.\ref{fig12}(a) and Fig.\ref{fig13}(a) exhibit similar patterns, where the nearest neighbors are predominantly located lateral to the pedestrians’ forward direction, which coheres with the findings of \citet{porzycki2017velocity} and \citet{cao2021spatial}. The spatial distribution map of nearest neighbors suggests that pedestrians adopt a crossing strategy that maintains substantial headway to safeguard velocity stability. As a result, pedestrians strive to eschew nearest neighbors positioned directly ahead during the crossing process, engendering
the distinctive spatial distribution observed. This phenomenon is markedly evident in experiments involving static context and less so
in dynamic context, implying a heightened spatial compression among pedestrians crossing dynamic crowds.

From Fig.\ref{fig12}(b) and Fig.\ref{fig13}(b), with the increase in global density, the decline of NNRD in both static and dynamic contexts has
presented a decaying trend, indicating the crossing strategy driven by low spatial constrictions. The lateral comparison shows that at
low global densities, the decline rate of NNRD in the dynamic context surpasses that in the static context, indicating cross pedestrians
experienced more spatial compression in dynamic context.

In the context of crowd cross, pedestrians engaging in bypassing maneuvers exhibit distinctive patterns. Specifically, individuals
opting to bypass from the right lead their movement along the left, a phenomenon corroborated by several studies \citep{moussaid2009experimental} and presented in the trajectory diagram. Such patterns suggest that pedestrians bypassing from the right are more likely to be in
close proximity to those on their left. Therefore, by assessing the positions of pedestrians’ nearest neighbors, detouring preferences can
be inferred.

\begin{figure}[ht!]
 \centering         
 \includegraphics[scale=0.85]{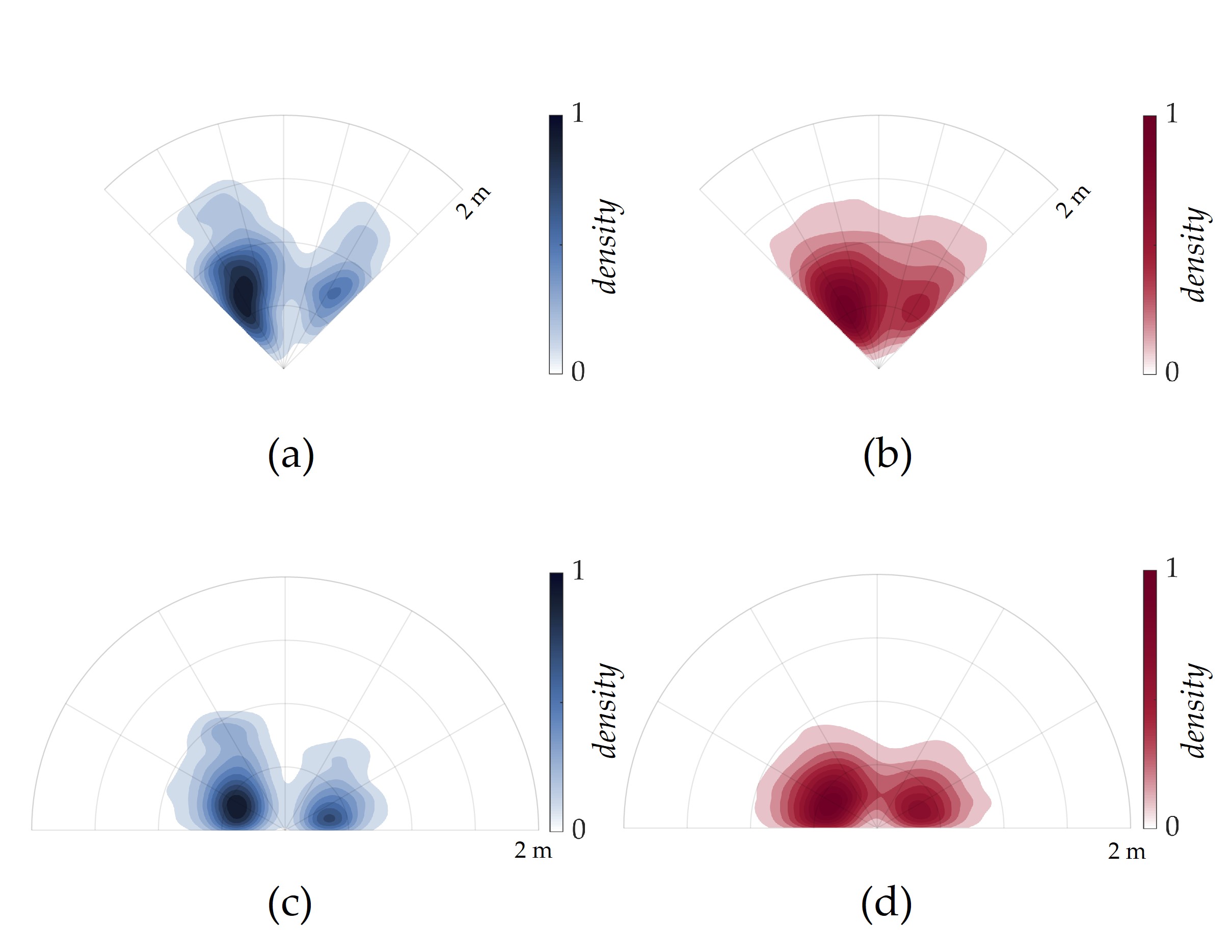}
\caption{Kernel density distributions of nearest neighbors for cross pedestrians in both dynamic and static crowds when $\phi$=45 degrees (above) and
$\phi$=90 degrees (below), subfigures (a), (c), and (b), (d) respectively depict the kernel density distributions of nearest neighbors for cross pedestrians in static and dynamic settings.
\label{fig14}}
\end{figure}

Fig.\ref{fig14} delineates the kernel density distribution of the nearest neighbors within the HFA for cross pedestrians in both
crowd patterns. The distribution of nearest neighbors to the left of the cross pedestrians is significantly more pronounced than that on the right. The uneven distribution trend is especially observable in the static crowd experiments. This trend persists, even when each
experiment group is scrutinized independently, as exemplified in Fig.\ref{fig12}(a) and Fig.\ref{fig13}(a). These patterns suggest a proclivity for
rightward movement among pedestrians during crossings, with this rightward bypassing tendency being particularly pronounced in
static crowd crossing. This concurs with the cultural norms prevalent in China, where pedestrians habitually walk on the right side.
These insights verified the preceding inference in section \ref{subsection6.2}. In static context, cross pedestrians tend to evade high-density areas,
particularly avoiding pedestrians ahead to evade deceleration, inducing the formation of cross-channel. Similar movement strategy
driven by low spatial constraints has been observed in relevant research, whereby crowds are observed to disperse ahead of the
intruder and converge behind the intruder \citep{nicolas2019mechanical,kleinmeier2020agent}.

\centerline{}

\subsection{Statistic analysis in detour mechanism}\label{subsection7.2}

\subsubsection{Deviation angle}\label{subsection7.2.1}

1 Deviation angle versus NNRA and NNRD

Certain factors contributing to pedestrian detour behaviors were identified in the microscopic evaluation of pedestrian movement
patterns. Two primary objective elements are commonly posited as instigators of pedestrian detours: directional conflicts and spatial
constraints. Detour behavior can be conceptualized as pedestrians opting to deviate from their initial path to mitigate potential clashes.
The NNRA can be employed to quantify the positional conflicts encountered by pedestrians, while the NNRD can serve as a metric to assess the spatial restrictions faced during pedestrian movements. Comprehensive definitions of the NNRA and NNRD are provided in
Section \ref{section4}. Pedestrians’ deviation angle is a quantitative index of their detour behaviors. An enlarged deviation angle signifies a heightened propensity to detour, and when the angle surpasses 90 degrees, it is indicative of a backward detour or spiral turning behavior.

\begin{figure}[ht!]
\plotone{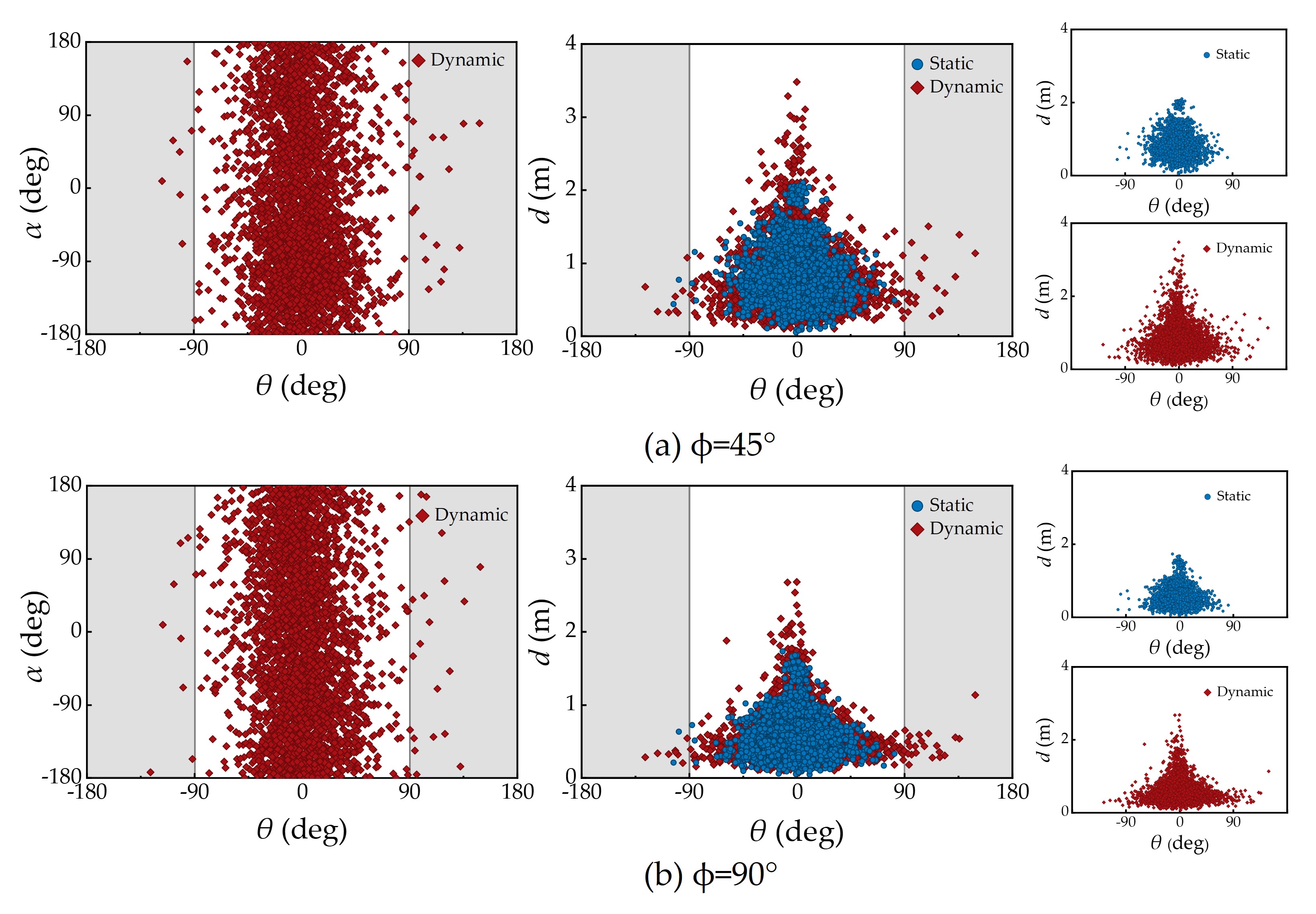}
\caption{Left: the relationship between the NNRA of cross pedestrians versus the deviation angle, right: the relationship between the NNRD of cross
pedestrians versus the deviation angle. The gray region in the figure represents the range of deviation angles greater than 90 degrees, and the data points in this region indicate pedestrians engaged in backward detour movements. (a) $\phi$= 45 degrees, (b) $\phi$= 90 degrees.
\label{fig15}}
\end{figure}

Statistical scrutiny of variations in pertinent parameters during pedestrian movements within static and dynamic crowds can afford
insights into factors modulating pedestrian detour behavior. Fig.\ref{fig15} exhibits scatter plots depicting the relationship between pedestrians’
NNRA and NNRD, corresponding to the deviation angle, under distinct eccentricity attentional angle conditions. In the scatter
plot correlating the NNRA with the deviation angle, no clear variation tendency is discernible among data points within the range of
the NNRA as the deviation angle augments. Conversely, in the scatter plot correlating the NNRD with the deviation angle, a decline in
the NNRD accompanies an increase in the deviation angle, and a triangular distribution is exhibited in the plot.

These findings suggest that spatial constraints significantly influence cross pedestrians during their movement, leading to a
preference for detours. Meanwhile, the influence of directional conflicts on pedestrian detour behavior appears comparatively limited.

\centerline{}

2 Deviation angle versus instantaneous velocity

\begin{figure}[ht!]
\plotone{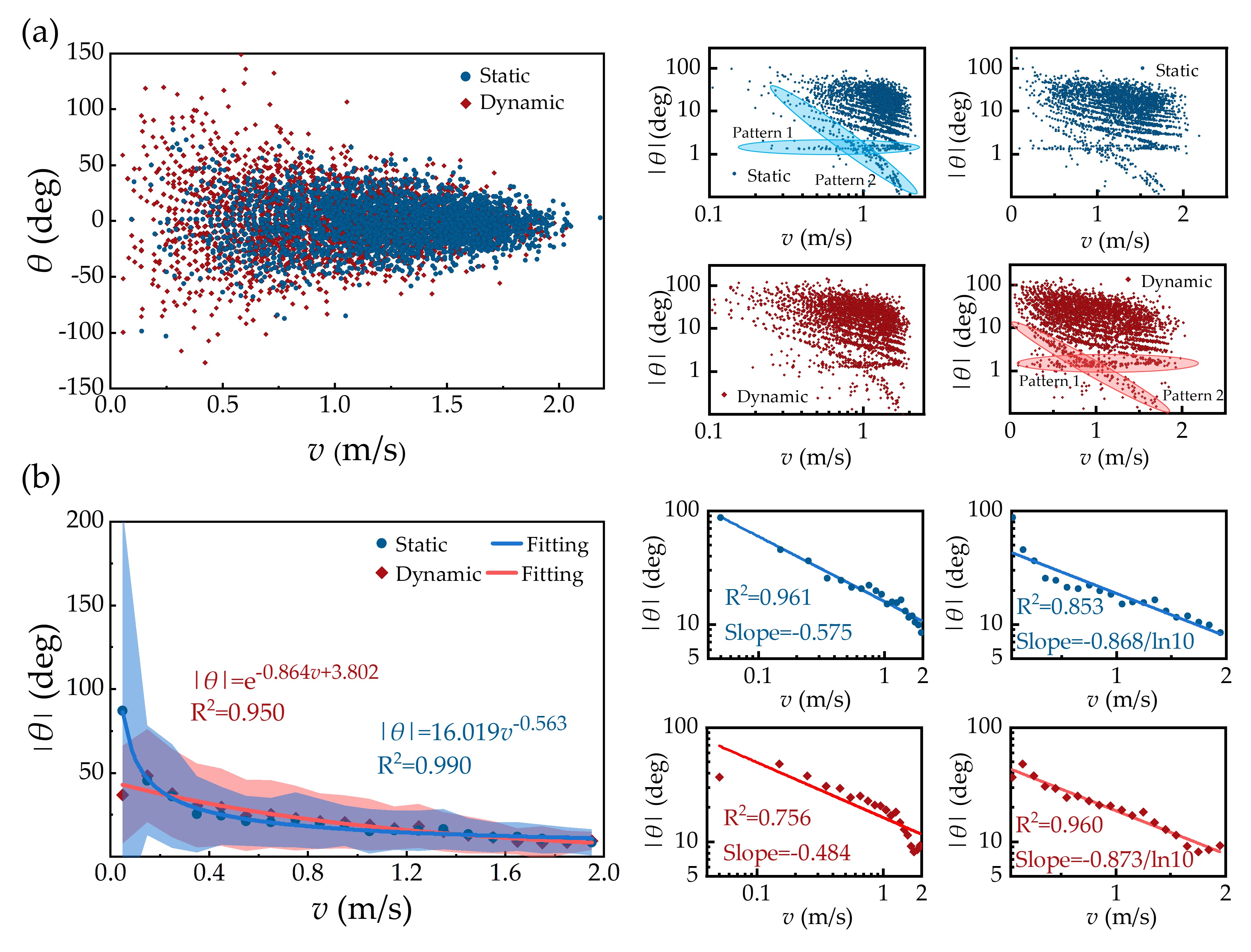}
\caption{(a) Scatter distribution relationship of deviation angle versus instantaneous velocity, the absolute value of deviation angle versus instantaneous
velocity: the dots indicate the mean value, and the shaded area represents a one-sigma error (mean ± standard deviation). The double logarithmic
representation reveals the power-law relationship between the deviation angle and instantaneous velocity, and the semi-logarithmic
representation reveals the exponential relationship between the deviation angle and instantaneous velocity.
\label{fig16}}
\end{figure}

We embarked on a statistic to scrutinize the relationship between pedestrian deviation angle and instantaneous velocity, the
outcomes of which are graphically presented in Fig.\ref{fig16}. The scatter plot vividly delineates the variation in the pedestrian deviation
angle contingent upon the instantaneous velocity. A noticeable observation is an inverse correlation between the pedestrian deviation
angle and the instantaneous velocity, which diminishes as the latter increases.

More specifically, for instantaneous velocity less than 0.5 m/s, the distribution of pedestrian deviation angles indicates a proclivity
for spreading along the vertical axis. A power or exponential decay function describes the relationship between the deviation angle and
the instantaneous velocity. During the dynamic experiments, the decay rate of the pedestrian deviation angle markedly accelerates
with an increase in speed, adhering to an exponential relationship, as elucidated in Fig.\ref{fig16}(b). The static experiments, however,
manifest a decay rate that conforms to a power-law relationship, as depicted in Fig.\ref{fig16}(a). The different decay trend of the deviation
angle indicates that pedestrians undergo more abrupt turning (faster angular changes) during detours within dynamic context.

We discern two distinct pedestrian movement patterns by analyzing the double-logarithmic and semi-logarithmic coordinate systems. Pattern 1 symbolizes the deceleration exhibited by pedestrians encountering obstacles, during which the deviation angle
remains relatively stable at around 0 degrees and is uninfluenced by fluctuations in instantaneous velocity. Contrastingly, Pattern 2
encapsulates the detour (along with decelerate) adopted by pedestrians upon confronting obstacles, typified by a deviation angle that
adheres to either a power law or an exponential decline as velocity amplifies. These two patterns collectively signify the distinct
pedestrian behavior during the crossing process, similar to the findings of \citet{parisi2016experimental}. In their research, pedestrians employ
three avoidance maneuvers to evade collisions: stoping, steering, stoping and steering.

While such findings might intuitively make sense, it is essential to highlight potential inaccuracies in data interpretation due to
certain movement mechanic issues. That is to say. In the context of static cross experiments, there is less observable deceleration
behavior exhibited by pedestrians encountering obstacles. Nevertheless, the presence of Pattern 1 was still discerned within logarithmic
coordinates, which could plausibly be ascribed to pedestrians’ anticipatory deceleration prior to circumventing obstacles.
Therefore, one might question whether this observed pattern could contribute to delineating two distinct patterns within the dynamic
context. A closer examination of Fig.\ref{fig16}(b) reveals additional evidence substantiating the existence of these dual movement modalities
in dynamic crowds. More specifically, in the lower velocity range, the mean value of the deviation angle in dynamic experiments
significantly undercuts that of the static counterpart.

\subsubsection{Angular velocity}\label{subsection7.2.2}

\begin{figure}[ht!]
\plotone{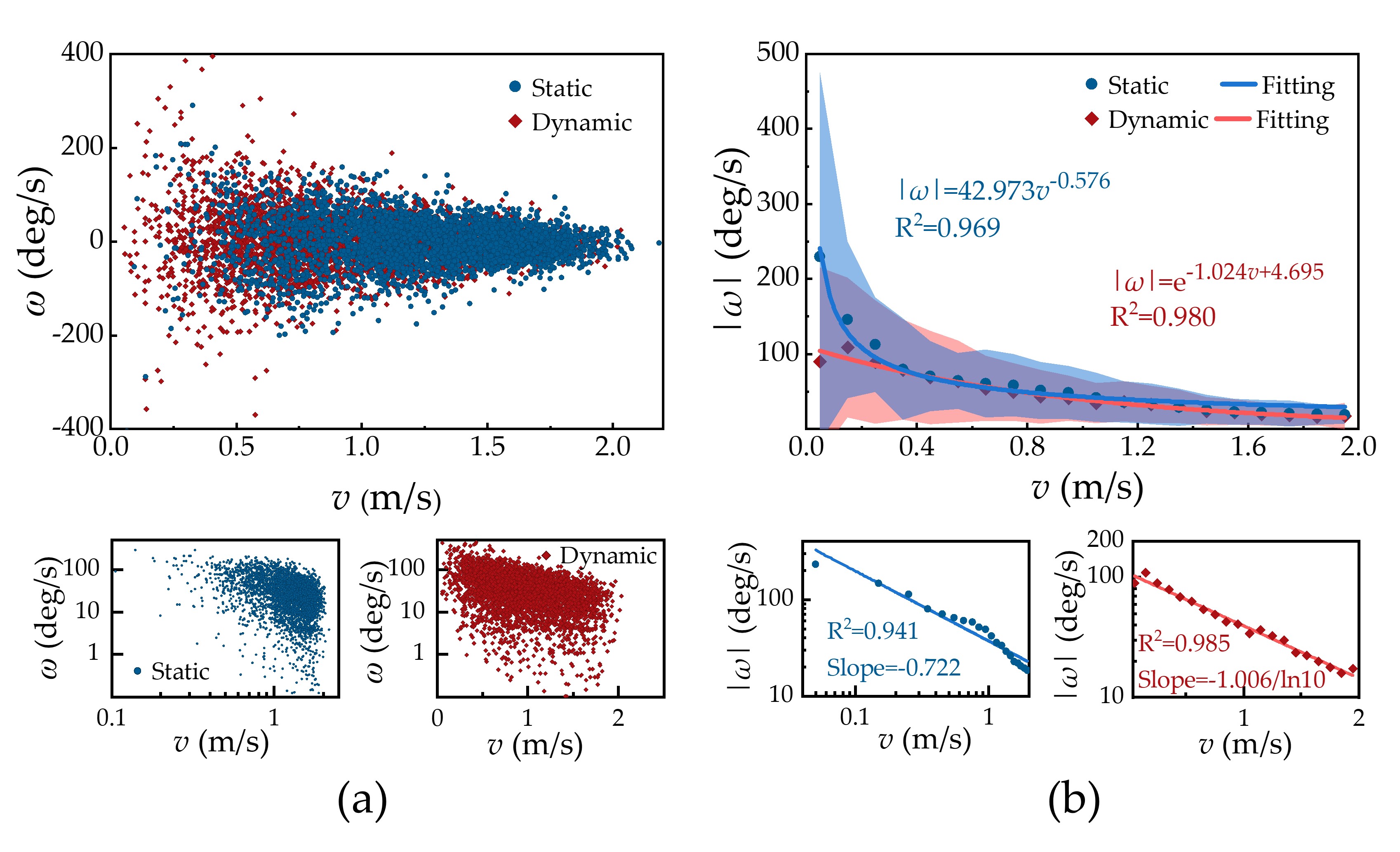}
\caption{(a) Scatter distribution relationship of angular velocity versus instantaneous velocity, the absolute value of angle velocity versus instantaneous velocity: the dots indicate the mean value, and the shaded area represents a one-sigma error (mean ± standard deviation). The double logarithmic representation reveals the power-law relationship between angular velocity and instantaneous velocity, the semi-logarithmic representation reveals the exponential relationship between angular velocity and instantaneous velocity.
\label{fig17}}
\end{figure}

Fig.\ref{fig17} illuminates the association between angular velocity and instantaneous velocity. The fitting results underscore a power-law
relationship for the decay of angular velocity in static scenarios, whereas, in dynamic contexts, the decay of angular velocity as velocity
escalates follows an exponential relationship, bearing similarity to the findings reported in the study by \citet{hicheur2005velocity}.

Consequently, with increased pedestrian speed, the angular velocity attenuates more rapidly when negotiating a dynamic crowd than a
static crowd. The statistical representation of angular velocity as a function of speed, as displayed in Fig. \ref{fig17}(b). The most pronounced
shifts in pedestrian angles (manifests as a large angular velocity) tend to occur during detouring or pathfinding. When we compare the change of angular velocity in different contexts, we find that in the lower speed ranges, the average angular
velocity in the dynamic experiment is significantly lower than in the static experiment. This observation suggests that some pedestrians
refrained from executing the detour maneuver. Such a finding lends further credence to the dual motion patterns postulated within
dynamic crowds.

\centerline{}
\centerline{}
\centerline{}

\centerline{}

\section{Conclusions} \label{section8}

To explore the reasons behind the persistence of crowd dynamism. This study undertook a pedestrian crossing experiment in two
crowd contexts: static and dynamic. Following this, a rigorous examination of the data was undertaken through the application of both macroscopic and microscopic analytical methodologies. All inferences and analyses within this study were predicated on the data
harvested from these experiments. The main issues addressed in this study are as follows:

1 The existence of persistent dynamism in crowds has been established. Analysis based on NNRD and velocity indicates that pedestrians
maintain dynamism in a highly constrained context.

2 Pedestrians exhibit different cross features in static and dynamic contexts. In the static context, the formation of crossing channels
is observed as a self-organizing pattern, while in the dynamic context, pedestrians exhibit unique behaviors such as backward detours
and spiral turning. Despite these, quantitative analysis results suggest that pedestrians adopt a consistent strategy in different crowd
contexts: detouring to overcome spatial limitations.

3 A statistical evaluation of deviation angles unveiled power-law and exponential decay trends corresponding to pedestrian deviation
angles vis-`a-vis velocity in static and dynamic crowds. Within the context of double-logarithmic and semi-logarithmic coordinate
systems, two pedestrian crossing patterns within the dynamic crowds were revealed: Pattern 1, demonstrating pedestrians
opting to decelerate upon encountering pedestrians, and Pattern 2, showing pedestrians choosing to detour upon confronting
pedestrians.

4 The findings addressed the questions proposed in this paper: pedestrians detouring causes crowds’ persistent collective dynamism.
The propensity of pedestrians to opt for detours to evade spatial constraints has been comprehensively observed within crowds.
Even amidst high-density human assemblies, certain individuals persistently exert attempts by seeking alternative pathways via
motion behaviors such as fallback and detour, irrespective of the potential for non-improved outcomes (designated as “aggressive
pedestrians”). Furthermore, because of the strategy to overcome space constraints, even minor disturbances can also impact the overall
dynamics of the crowd. The “catfish effect” engendered by these aggressive pedestrians may play a significant role in forming
turbulence.

This study will provide empirical support for crowd dynamic and pedestrian trajectory prediction research.

\section{Data availability} \label{section9}
The data can be found here: 
 {\url{https://drive.google.com/drive/folders/1NYVnRp0z8VPuskfezMr51gB-sraOf6Iq?usp=drive_link}}

\bibliographystyle{aasjournal}



\end{document}